\pdfoutput=1
\def\COMBINEDVERSION{1}
\documentclass[11pt]{article}

\usepackage[margin=1in]{geometry}
\usepackage[T1]{fontenc}
\usepackage{lmodern}
\usepackage{setspace}
\usepackage{booktabs}
\usepackage{tabularx}
\usepackage{array}
\usepackage{longtable}
\usepackage{enumitem}
\usepackage{amsmath,amssymb}
\usepackage{graphicx}
\usepackage{float}
\usepackage{placeins}
\usepackage{xcolor}
\usepackage{tikz}
\usepackage[authoryear,round,sort]{natbib}
\usepackage{url}
\usepackage{titlesec}
\usepackage{fancyhdr}
\usepackage{hyperref}

\definecolor{navy}{HTML}{173B57}
\definecolor{bluefill}{HTML}{E8F1F7}
\definecolor{teal}{HTML}{257A76}
\definecolor{tealfill}{HTML}{E5F3F1}
\definecolor{redaccent}{HTML}{A7423E}
\definecolor{redfill}{HTML}{F8E9E7}
\definecolor{goldaccent}{HTML}{A8781D}
\definecolor{goldfill}{HTML}{FAF2D9}
\definecolor{graytext}{HTML}{4E5963}
\definecolor{grayline}{HTML}{AAB4BC}
\definecolor{grayfill}{HTML}{F3F5F6}
\definecolor{strongfill}{HTML}{9FC5AE}
\definecolor{emergingfill}{HTML}{E5D39F}
\definecolor{thinfill}{HTML}{F3E3E2}

\hypersetup{
  colorlinks=true,
  linkcolor=navy,
  citecolor=teal,
  urlcolor=redaccent,
  pdftitle={From Metrics to Decisions in NBA Analytics: A Critical Integrative Review and Decision-Readiness Framework},
  pdfauthor={Yang Zhou and Tianyu Guan}
}

\setlist[itemize]{leftmargin=1.4em,itemsep=0.2em,topsep=0.35em}
\setlist[enumerate]{leftmargin=1.6em,itemsep=0.2em,topsep=0.35em}
\titleformat{\section}{\Large\bfseries\color{navy}}{\thesection}{0.7em}{}
\titleformat{\subsection}{\large\bfseries\color{navy}}{\thesubsection}{0.6em}{}
\titleformat{\subsubsection}{\normalsize\bfseries\color{graytext}}{\thesubsubsection}{0.5em}{}
\newcolumntype{Y}{>{\raggedright\arraybackslash}X}
\newcolumntype{P}[1]{>{\raggedright\arraybackslash}p{#1}}

\title{\vspace{-1.5em}\textbf{From Metrics to Decisions in NBA Analytics: A Critical Integrative Review and Decision-Readiness Framework}}
\author{%
  Yang Zhou$^{1}$ \qquad Tianyu Guan$^{2,1,*}$\\[0.6em]
  \small $^{1}$TY SPORTS ANALYTICS INC., Canada\\
  \small $^{2}$Department of Mathematics and Statistics, York University,\\
  \small Toronto, Canada\\[0.45em]
  \small Yang Zhou: \href{mailto:yangzhou@tysportsanalytics.com}{yangzhou@tysportsanalytics.com}\\
  \small $^{*}$Corresponding author: Tianyu Guan, \href{mailto:tguan@yorku.ca}{tguan@yorku.ca}
}
\date{}

\begin{document}

\maketitle

\begin{abstract}
National Basketball Association (NBA) teams have increasingly detailed metrics, but better predictions do not necessarily improve decisions. This critical integrative review draws on prior reviews, citation tracing, and topic searches across seven research streams: on-court action, player value, role, lineup synergy, availability, draft and development, and contracts and roster construction. An observation--state--action--decision--evaluation chain organizes the synthesis. Six decision-readiness gates guide our assessment: point-in-time validity, uncertainty, context portability, action feasibility, opportunity-set observability, and evaluation, with requirements matched to each claim. The reviewed literature is strongest in measuring and predicting individual components of a decision. Evidence is less developed at interfaces that combine components, transfer them across settings, and compare feasible actions. We outline a proposed deployment workflow, a reporting contract, and a research agenda covering player transport, role substitution, roster fragility, legal action generation, and asset valuation. The 2023 collective bargaining agreement and forthcoming 3-2-1 Draft Lottery illustrate how institutional changes generate research questions. Models should inform evaluable comparisons of feasible choices. While its effect on organizational decision quality remains an empirical question, the framework provides a diagnostic and reporting structure for matching decision claims to evidence requirements.
\end{abstract}


\section{Introduction}

NBA analytics now reaches well beyond box-score comparisons. Play-by-play records support possession-level analysis, lineup data describe combinations of players, and optical tracking records movement on and off the ball. Schedules, injury reports, contracts, transactions, and prospect records extend this work to decisions about participation and team building. Researchers have used these sources to estimate player contribution, value possessions, describe roles, forecast injuries and draft outcomes, and construct rosters. Prior reviews and foundational work document the statistical foundations, performance measures, and data sources on which these applications depend \citep{Kubatko2007,Sarlis2020,Terner2021,Huyghe2022,Kovalchik2023,ZhouLi2024,Chen2025}.

These works approach the field from several directions. Kubatko et al. establish possession-based statistical foundations; Sarlis and Tjortjis survey player and team evaluation; and Terner and Franks examine models of game play, strategy, and performance, identifying causal inference as a major research need. Huyghe et al. focus on the player and contextual constraints affecting NBA game play. Kovalchik covers tracking methods and action valuation across sports, while Zhou and Li organize basketball research by dimension, granularity, task, and stakeholder question. Chen et al. assess data--method alignment across 206 performance-analysis studies. Table~\ref{tab:priorreviews} sets out how our review builds on this work. We take a particular decision---who acts, when, and with which alternatives---as the starting point, then ask whether the available estimates can support it. This perspective connects on-court research with availability, development, draft, contract, transaction, and roster decisions.

\begin{table}[htbp]
\centering
\caption{Positioning this review relative to prior reviews and foundational basketball work.}
\label{tab:priorreviews}
\footnotesize
\begin{tabularx}{\textwidth}{P{2.65cm}P{3.1cm}Y}
\toprule
\textbf{Source} & \textbf{Primary organizing lens} & \textbf{Boundary addressed by the present review} \\
\midrule
\citet{Kubatko2007} & Possession accounting and foundational basketball statistics & Connects statistical measurement to specific coaching and personnel actions \\
\citet{Sarlis2020} & Methods and metrics for player and team evaluation & Separates observed performance, portable ability, price, and team-specific decision value \\
\citet{Terner2021} & Statistical models of player and team performance & Extends from estimators to action feasibility, organizational context, and deployment evaluation \\
\citet{Huyghe2022} & Player and contextual constraints underpinning NBA game play & Extends beyond game-play determinants to development, draft, contracts, and roster choices \\
\citet{Kovalchik2023} & Player-tracking data and methods across sports & Examines whether tracking outputs support interventions and evaluated NBA decisions \\
\citet{ZhouLi2024} & Dimensions, granularity, tasks, and stakeholder questions in professional basketball & Adds a decision timestamp, feasible and plausibly available actions, coupled cross-stream states, and explicit evaluation \\
\citet{Chen2025} & Data--method alignment in basketball performance analysis & Shifts to claim-specific time validity, uncertainty, transport, feasible and available actions, and evaluation \\
\textbf{Present review} & Observation--state--action--decision--evaluation chain & Connects seven operations streams and sets out research and reporting priorities \\
\bottomrule
\end{tabularx}
\end{table}

The difficulty becomes clear when a model leaves the setting in which it was estimated. A plus-minus coefficient may summarize contribution alongside observed teammates yet say little about performance in a new role. A lineup forecast may rank five-player units without accounting for fatigue or the bench minutes needed to complete a rotation. Similarly, predicting injury risk does not identify the effect of resting a player, and predicting salary does not establish what a contract is worth to a particular team. These are gaps between the question a model answers and the choice its user faces. More granular data or a more complex algorithm may help, but neither resolves the mismatch on its own.

Our concern is how estimates enter basketball-operations decisions. Such decisions concern a specific team at a specific time, use incomplete information, and draw on alternatives that may be difficult to observe. A player's projected contribution matters alongside the role available to that player, the price of acquiring the contract, and the team's tolerance for risk. The analytical target is the value of an action under those conditions, over an appropriate horizon.

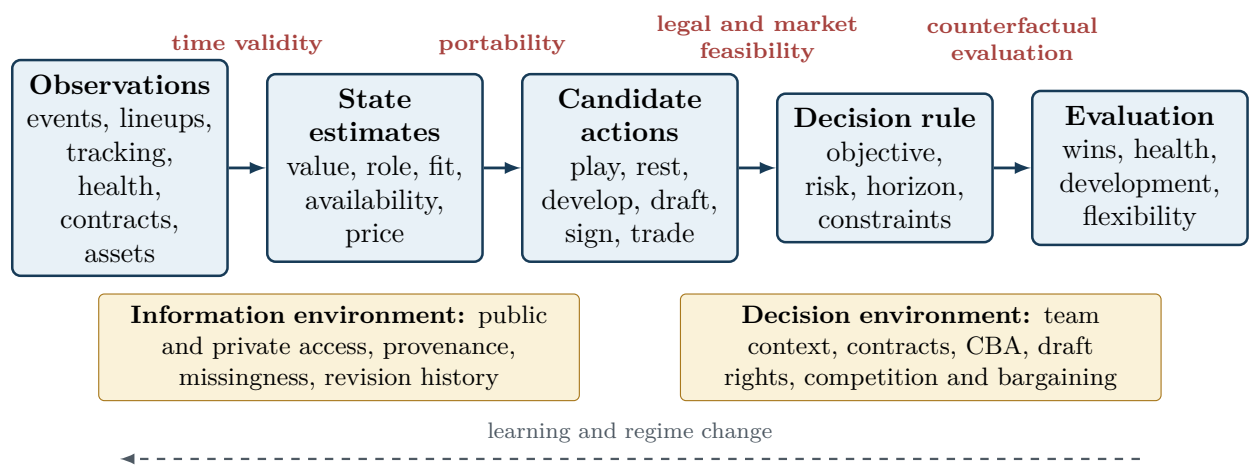
\begin{figure}[htbp]
  \centering
  \resizebox{\textwidth}{!}{\begin{tikzpicture}[
  every node/.append style={execute at begin node={\hyphenpenalty=10000\relax\exhyphenpenalty=10000\relax}},
  font=\small,
  stage/.style={rounded corners=3pt, draw=navy, line width=0.9pt, fill=bluefill,
                minimum height=1.45cm, text width=2.35cm, align=center, inner sep=5pt},
  gate/.style={font=\scriptsize\bfseries, text=redaccent, align=center},
  flow/.style={->, >=latex, line width=1.0pt, draw=navy},
  feedback/.style={->, >=latex, line width=0.8pt, draw=graytext, dashed},
  context/.style={rounded corners=2pt, draw=goldaccent, fill=goldfill,
                  text width=5.75cm, minimum height=0.9cm, align=center,
                  inner sep=4pt, font=\footnotesize}
]
  \node[stage] (obs) at (-6.4,0) {\textbf{Observations}\\events, lineups, tracking, health, contracts, assets};
  \node[stage] (state) at (-3.2,0) {\textbf{State estimates}\\value, role, fit, availability, price};
  \node[stage] (actions) at (0,0) {\textbf{Candidate actions}\\play, rest, develop, draft, sign, trade};
  \node[stage] (choice) at (3.2,0) {\textbf{Decision rule}\\objective, risk, horizon, constraints};
  \node[stage] (outcome) at (6.4,0) {\textbf{Evaluation}\\wins, health, development, flexibility};

  \draw[flow] (obs) -- (state);
  \draw[flow] (state) -- (actions);
  \draw[flow] (actions) -- (choice);
  \draw[flow] (choice) -- (outcome);
  \node[gate] at (-4.8,1.55) {time validity};
  \node[gate] at (-1.6,1.55) {portability};
  \node[gate] at (1.6,1.62) {legal and market\\feasibility};
  \node[gate] at (4.8,1.62) {counterfactual\\evaluation};

  \node[context] (info) at (-3.65,-2.25)
    {\textbf{Information environment:} public and private access, provenance, missingness, revision history};
  \node[context] (rules) at (3.65,-2.25)
    {\textbf{Decision environment:} team context, contracts, CBA, draft rights, competition and bargaining};

  \draw[feedback] (6.4,-3.65) --
    node[midway, above=2pt, font=\scriptsize, text=graytext] {learning and regime change}
    (-6.4,-3.65);
\end{tikzpicture}}
  \caption{The NBA analytics data-to-decision pipeline. Model performance at one stage does not validate the next interface. Decision-ready evidence must preserve the information available at the decision time, transport estimates to the intended context, generate feasible actions, and evaluate choices against credible alternatives.}
  \label{fig:pipeline}
\end{figure}

We organize the review around the evidence needed for these choices. Contribution, role, synergy, and availability describe related aspects of a player's value: what the player can produce, in which setting, alongside whom, and at what workload. Contracts, draft rights, and the collective bargaining agreement (CBA) shape the cost and feasibility of acting on those estimates. Figure~\ref{fig:pipeline} shows the links from observations to an evaluated decision.

The review makes three connected contributions. It synthesizes evidence across seven NBA research streams to identify where well-studied components remain weakly connected to decisions. It integrates established methodological requirements into six decision-readiness gates, using the dated team decision as the organizing unit and matching the requirements to each claim level. It then translates these diagnoses into a reporting contract and proposed deployment workflow (Section~\ref{sec:team-decisions}). The contribution lies in this integration and operationalization, rather than in a new statistical estimator. The 2023 CBA and the 3-2-1 Draft Lottery approved in 2026 illustrate how changes in constraints and incentives generate research questions.

The scope covers competitive basketball operations: on-court action, player evaluation and development, availability, draft, contracts, transactions, and roster construction. Betting, fan sentiment, ticket demand, media value, and officiating fall outside it unless they directly affect one of these decisions. Throughout, we assess a study against the claim it makes. Descriptive research has value in its own right; the additional requirements arise when its findings are used to support a choice.

\section{Review Purpose, Scope, and Analytical Lens}

\subsection{The questions that organize the review}

The review is organized around three questions.

\begin{enumerate}
  \item \textbf{What decision-relevant states does NBA research estimate?} This includes player contribution, role and capability, lineup synergy, availability and workload capacity, development, market value, and contract value.
  \item \textbf{How far does each research stream travel from observation to action?} We distinguish measurement, prediction, causal or counterfactual estimation, constrained prescription, and prospective evaluation.
  \item \textbf{Which interfaces must be built next?} We examine connections among player states, team context, legal constraints, market opportunities, uncertainty, and evaluation, with particular attention to the new CBA and lottery regimes.
\end{enumerate}

These questions guide the selection and comparison of studies. We give particular attention to work that changes the target of analysis, strengthens validation, measures a previously unobserved state, or makes a decision amenable to evaluation. Small differences in model performance receive less attention when they leave the underlying decision claim unchanged. Selection prioritizes evidence that directly informs the decision interfaces under review, rather than comprehensive coverage of each field.

\subsection{Review approach and evidence selection}

A critical integrative review suits this purpose because the relevant studies differ in their units of analysis, terminology, data access, and publication format. We use the evidence map, available as a \href{https://github.com/yangzhou-tysportsanalytics/From-Metrics-to-Decisions-in-NBA-Analytics/blob/main/nba-decision-readiness-evidence-map.csv}{machine-readable CSV on GitHub}, to examine how their findings connect and how far their claims extend. The review is not an exhaustive census or a PRISMA-style systematic review, and its study counts cannot be used to estimate the prevalence of methods or findings in NBA research.

The evidence cutoff was 1 September 2026. We began with the terminology and seed references in Table~\ref{tab:priorreviews}, then followed backward and forward citations into statistics, computer science, sports medicine, economics, operations research, and management. Targeted searches supplemented this process. Sources included scholarly web search and publisher or DOI pages; PubMed/MEDLINE and PubMed Central for health and availability; arXiv, PMLR, SSRN, MIT Sloan proceedings, institutional repositories, and author pages for methodological work; and official NBA and NBA Players Association archives for rule states. Search terms combined basketball or NBA with performance and plus-minus; tracking, spacing, defense, and action value; role, lineup, rotation, and network; injury, workload, and participation; or draft, development, salary, contract, trade, and roster optimization. We followed further leads when a paper introduced a relevant estimand, data layer, validation design, or decision interface. Search functions differed across sources, so the searches cannot be treated as equivalent database runs. The companion materials record the query families, access status, and retained evidence.

Eligible sources included primary empirical studies, methodological papers demonstrated on NBA data, and full conference papers. A source had to estimate a decision-relevant state, examine transport or uncertainty, model an intervention or constrained action, or document a relevant data or institutional limitation. Broader basketball and multisport studies were included where they directly informed an NBA decision interface. Reviews supplied context and terminology. Official NBA and NBA Players Association documents established legal and policy states; they were not used as evidence of a rule's effects.

Five discovery workstreams produced 132 records. These were search-management categories; the synthesis subsequently grouped the evidence into seven substantive research streams. Reconciliation by DOI, normalized title, and report version left 127 unique reports, of which 51 were retained. Seven prior-review and NBA foundations brought the NBA synthesis corpus to 58 reports: 50 inspected in full text, five available through abstracts or verified metadata and used only for positioning or scope claims, and three official rule documents. Eight non-NBA methodological works support the framework separately and do not enter the NBA evidence counts or maturity judgments. The supplementary search-and-synthesis record documents eligibility, deduplication, counts, and the rationale for every cell of Figure~\ref{fig:gapmap}. A 66-row evidence map, available as a \href{https://github.com/yangzhou-tysportsanalytics/From-Metrics-to-Decisions-in-NBA-Analytics/blob/main/nba-decision-readiness-evidence-map.csv}{machine-readable CSV on GitHub}, links all cited sources to their roles in the review. Supplementary Table~S4 provides the report-level crosswalk of claim levels, decision interfaces, and principal limitations.

For each retained report, we extracted the observation unit, data access, target or estimand, validation or identification design, uncertainty output, and claimed decision. We then assessed what further evidence would be needed to use the estimate in that decision. The extracted evidence dimensions were assessed separately rather than combined into a quality score. Public reproducibility, for example, says little on its own about a model's relevance to a trade, while restricted medical data may support an operationally useful study. Figure~\ref{fig:gapmap} compares the maturity of research interfaces on this basis; it does not rank papers or count publications. Both authors worked jointly on screening, extraction, and coding. During manuscript preparation, both authors independently cross-checked the source claims and the extracted information for all 66 records included in Table~S4, as well as all Figure 4 cell ratings. Discrepancies were resolved through joint review, and the final coding was confirmed by both authors. No formal estimate of inter-rater agreement is reported.

\subsection{Claim levels and decision readiness}

We distinguish five claim levels:

\begin{enumerate}
  \item \textbf{Measurement:} What happened, and how should observed contribution or behavior be summarized?
  \item \textbf{Prediction:} What outcomes can be predicted in a declared future or held-out sample, using a specified information set?
  \item \textbf{Causal or counterfactual estimation:} What would change under a specified intervention or alternative context?
  \item \textbf{Constrained prescription:} Which action is preferred under an explicit objective, information set, and feasible action set?
  \item \textbf{Prospective evaluation:} Did an implemented decision rule improve outcomes under deployment against a credible comparator?
\end{enumerate}

Retrospective replay can support a decision claim under stated identification and support assumptions, but does not constitute prospective deployment evaluation. Scientific credibility concerns whether a study answers its stated question; decision readiness concerns whether its output can inform a choice. We assess the latter through six decision-readiness gates: point-in-time validity, uncertainty, context portability, action feasibility, opportunity-set observability, and evaluation. The requirements depend on the claim. A descriptive paper has no obligation to enumerate legal trades, but a trade recommendation must address both legality and counterpart acceptance. Constrained prescription names a claim level; action feasibility names one gate used to assess it. Public reproducibility and organizational auditability are recorded separately, including for studies based on protected data.

\section{A Decision-Readiness Framework for NBA Analytics}

\subsection{Specifying the decision}

Let $I_t$ be the information available to a team at date $t$, $R_t$ the relevant rule and contract state, $M_t$ a specified scenario for the partly observed market opportunity set, and $\mathcal{A}(I_t,R_t,M_t)$ the actions admissible in that scenario. Let $Y_{t:t+h}(a)$ denote the multidimensional consequences of action $a$ over a common declared horizon $h$, and let $V_{t:t+h}(a)=U(Y_{t:t+h}(a))$ be their scalar organizational value under a declared objective. Define the corresponding scalar loss as $L_{t:t+h}(a)=-V_{t:t+h}(a)$. For transactions, $M_t$ depends jointly on each required party's reservation value, outside option, and latent willingness to transact; legal admissibility is therefore necessary but not sufficient for a transaction to be treated as plausibly available. A generic mean--risk decision is

\begin{equation}
  a_t^*(M_t) = \arg\max_{a \in \mathcal{A}(I_t,R_t,M_t)}
  \left\{ \mathbb{E}[V_{t:t+h}(a)\mid I_t,R_t,M_t] - \lambda\,\rho(L_{t:t+h}(a)\mid I_t,R_t,M_t) \right\}.
  \label{eq:decision}
\end{equation}

The dependence on $I_t$ and $R_t$, held fixed in each scenario comparison, is suppressed on the left-hand side.

The outcome vector can include wins, playoff advancement, player development, health, payroll, asset value, and future flexibility. The mapping $U$ converts those outcomes to a scalar value; it does not by itself specify the organization's full risk preference. Downside aversion is represented by $\lambda\geq 0$ and the risk functional $\rho$, applied to the conditional loss distribution. Examples include expected excess loss above a specified threshold and conditional value-at-risk under an explicit loss convention. CVaR at the 95th percentile, for instance, can summarize mean loss in the worst five percent of joint availability, playoff, and asset-value scenarios.

For decisions spanning several seasons, $U$ must specify how outcomes at different dates are discounted or weighted. Those weights may reflect the team's competitive timeline and mandate. The objective, time weights, and risk preferences are organizational inputs to be declared or elicited; predictive fit alone does not determine them. Transaction flexibility is included in $Y$, avoiding a second risk penalty for the same consequence. This formulation follows the distinction among actions, uncertain consequences, values, and losses in decision analysis and incorporates constraints in the feasible set \citep{Berger1985,BoydVandenberghe2004}.

For a causal or prescriptive claim, $Y(a)$ denotes an action-specific potential outcome. It cannot generally be recovered by averaging observed outcomes among cases that selected $a$. If $I_t^{\mathrm{obs}}$ is the information observed by the researcher and $D_t$ the realized action, then $\mathbb{E}[Y\mid I_t^{\mathrm{obs}},D_t=a]$ need not equal $\mathbb{E}[Y(a)\mid I_t^{\mathrm{obs}}]$. Unrecorded coaching, player, or organizational information may affect both selection and outcomes.

Equation~\ref{eq:decision} compares actions within a stated market scenario. A team may know terms that the researcher cannot observe. If acceptance is uncertain to the team itself, the controllable action is to submit an offer; acceptance, rejection, and negotiation costs then enter $Y(a)$. Beliefs about those outcomes must carry through expected value and downside risk. A recommendation to complete a transaction remains conditional on attainable terms. Many studies estimate contribution or injury risk; fewer identify action effects, reconstruct available opportunities, or evaluate the resulting choice.

Each part of the decision object motivates a readiness gate. The time index on $I_t$ establishes the need for point-in-time validity. Value and loss distributions, together with $\rho$, require uncertainty to be represented in the choice. Applying an outcome model to a new team, role, opponent, or rule regime raises the question of context portability. Membership in $\mathcal{A}$ requires legal and operational feasibility, while $M_t$ makes explicit the need to observe, reconstruct, or bound the available opportunities. Finally, comparing $U(Y)$ across actions calls for an evaluation design suited to the claim. The six decision-readiness gates thus identify different ways an otherwise credible estimate can fail to support a choice. Reproducibility and organizational auditability allow others to inspect that assessment.

Consider two models that rank all realistically available players in the same order. The more accurate model may leave a team's choice unchanged while improving its assessment of uncertainty or downside. An estimate of role portability or joint availability could be more useful if it changes which alternatives the team should consider. Value-of-information analysis formalizes this link between information, choice, and outcome \citep{Howard1966}. It helps explain why improvements in average predictive error can have limited operational value.

\subsection{Player value is a coupled state}

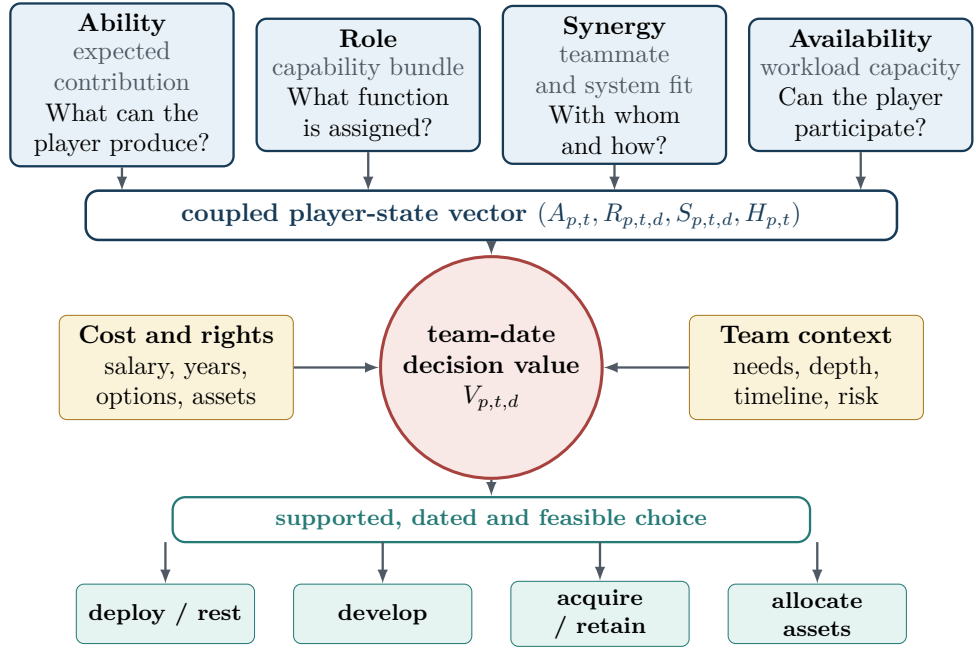
\begin{figure}[htbp]
  \centering
  \resizebox{0.78\textwidth}{!}{\begin{tikzpicture}[
  font=\small,
  state/.style={rounded corners=4pt, draw=navy, line width=0.9pt, fill=bluefill,
                text width=2.8cm, minimum height=1.72cm, align=center, inner sep=4pt},
  coupled/.style={rounded corners=5pt, draw=navy, line width=1.0pt, fill=white,
                  minimum width=11.2cm, minimum height=0.68cm, align=center,
                  font=\small\bfseries, text=navy},
  center/.style={circle, draw=redaccent, line width=1.2pt, fill=redfill,
                 minimum size=2.55cm, align=center, font=\small\bfseries},
  constraint/.style={rounded corners=3pt, draw=goldaccent, fill=goldfill,
                     text width=2.95cm, minimum height=1.05cm, align=center, inner sep=4pt},
  decisionhub/.style={rounded corners=5pt, draw=teal, line width=0.9pt, fill=white,
                      minimum width=8.8cm, minimum height=0.62cm, align=center,
                      font=\footnotesize\bfseries, text=teal},
  action/.style={rounded corners=3pt, draw=teal, fill=tealfill,
                 text width=2.25cm, minimum height=0.82cm, align=center,
                 inner sep=3pt, font=\footnotesize\bfseries},
  into/.style={->, >=latex, line width=0.9pt, draw=graytext}
]
  \node[state] (ability) at (-5.1,3.90)
    {\textbf{Ability}\\[-1pt]\textcolor{graytext}{expected contribution}\\What can the player produce?};
  \node[state] (role) at (-1.7,3.90)
    {\textbf{Role}\\[-1pt]\textcolor{graytext}{capability bundle}\\What function is assigned?};
  \node[state] (synergy) at (1.7,3.90)
    {\textbf{Synergy}\\[-1pt]\textcolor{graytext}{teammate and system fit}\\With whom and how?};
  \node[state] (avail) at (5.1,3.90)
    {\textbf{Availability}\\[-1pt]\textcolor{graytext}{workload capacity}\\Can the player participate?};

  \node[coupled] (statevec) at (0,2.10)
    {coupled player-state vector $\left(A_{p,t},R_{p,t,d},S_{p,t,d},H_{p,t}\right)$};
  \foreach \n/\x in {ability/-5.1,role/-1.7,synergy/1.7,avail/5.1} {
    \draw[into] (\n.south) -- (\x,2.44);
  }

  \node[center] (dv) at (0,0)
    {team-date\\decision value\\$V_{p,t,d}$};
  \draw[into] (statevec.south) -- (dv.north);

  \node[constraint] (cost) at (-4.35,0)
    {\textbf{Cost and rights}\\salary, years, options, assets};
  \node[constraint] (context) at (4.35,0)
    {\textbf{Team context}\\needs, depth, timeline, risk};
  \draw[into] (cost.east) -- (dv.west);
  \draw[into] (context.west) -- (dv.east);

  \node[decisionhub] (choice) at (0,-2.10) {supported, dated and feasible choice};
  \draw[into] (dv.south) -- (choice.north);

  \node[action] (deploy) at (-4.5,-3.40) {deploy / rest};
  \node[action] (develop) at (-1.5,-3.40) {develop};
  \node[action] (acquire) at (1.5,-3.40) {acquire / retain};
  \node[action] (assets) at (4.5,-3.40) {allocate assets};
  \foreach \n/\x in {deploy/-4.5,develop/-1.5,acquire/1.5,assets/4.5} {
    \draw[into] (\x,-2.41) -- (\n.north);
  }
\end{tikzpicture}}
  \caption{An integrated player-state model. Ability, role, synergy, and availability form a coupled player-state vector; cost, rights, and team context determine whether and how that state creates value for a particular team and dated decision. Arrows indicate dependencies, not a claim that one universal functional form is correct.}
  \label{fig:states}
\end{figure}

A single rating is convenient for comparing players, but a team needs to know how that value depends on the intended use. For player $p$ at time $t$ in decision context $d$, we write

\begin{equation}
  V_{p,t,d} = f\!\left(A_{p,t},\,R_{p,t,d},\,S_{p,t,d},\,H_{p,t},\,C_{p,t,d},\,Z_{t,d}\right),
  \label{eq:value}
\end{equation}

where $A$ is ability or expected contribution, $R$ is role, $S$ is synergy with teammates and system, $H$ is availability or workload capacity, $C$ is cost and contractual control, and $Z$ describes the team's competitive and institutional context. Relevant elements of $Z_{t,d}$ include playoff position, roster depth, competitive timeline, apron or hard-cap status, and the rules in force. We impose no additive or linear form on $f$. The value of a player's ability may depend on role, health, and contractual rights, with institutional thresholds changing the value of the same component across teams. Figure~\ref{fig:states} depicts these dependencies.

For example, a deep contender may be able to absorb a player's absences and offer a complementary role, whereas a shallow roster may face expensive replacement needs. The baseline contribution estimate alone would miss that difference. Cost, rights, and the team's competitive timeline determine whether the projected contribution justifies acquiring or retaining the player.

\paragraph{Illustrative decision trace.} Consider two general managers assessing the same high-impact wing at a hypothetical trade deadline $t$. One leads a deep contender; the other leads a top-heavy team near an apron threshold. Both use the same baseline contribution estimate, absence scenarios, and outcome horizon $h$. The candidate actions are the wing acquisition package ($A$), an alternative capability bundle ($B$), and retaining the current roster ($C$). Suppose baseline contribution favors $A$. Table~\ref{tab:decisiontrace} illustrates how a legal screen can change that comparison before role, risk, and market terms determine a recommendation.

\begin{table}[H]
\centering
\caption{Illustrative decision trace for two team contexts.}
\label{tab:decisiontrace}
\footnotesize
\renewcommand{\arraystretch}{1.08}
\begin{tabularx}{0.98\textwidth}{P{2.35cm}Y Y}
\toprule
\textbf{Readiness gate} & \textbf{Deep contender} & \textbf{Top-heavy apron team} \\
\midrule
Point-in-time validity & Freeze $I_t$, contracts, rights, objective, and horizon before comparing actions & Use the same cutoff and horizon, but freeze this team's own roster and rule state \\
Uncertainty & Assess whether replacement depth buffers lost minutes under the stated absence scenarios & Estimate downside under the same absence scenarios using this team's replacement capacity \\
Context portability & Assess whether projected contribution transports to the complementary role; require evidence for that deployment & Assess role overlap and test contribution across plausible role and minutes assignments \\
Action feasibility & In this hypothetical screen, $A$, $B$, and $C$ are legal; assess their rights and future-tool costs & Package $A$ requires a transaction tool unavailable to this team at $t$; exclude $A$ and compare legal actions $B$ and $C$ \\
Opportunity-set observability & Counterpart acceptance is unknown; condition acquisition on acceptable, attainable terms & Apply the same acceptance check to this team's affordable package set; legality alone is insufficient \\
Evaluation & Archive the action, forecast, and baseline; assess contribution, depth, and flexibility over $h$ & Preserve the same outcome definitions, but declare this team's weights and downside tolerance in advance \\
\bottomrule
\end{tabularx}
\end{table}
The screen leaves the contender comparing $A$, $B$, and $C$, while the apron team compares $B$ with $C$, despite the shared baseline ranking. The contender could prefer $A$ if its role-adjusted contribution outweighs cost and downside; the apron team could prefer $B$ only if the bundle improves on retention under its own objective. These are hypothetical comparisons, conditional on acceptable, attainable terms. Changes in price, absence risk, or replacement quality warrant reassessment. Evaluating either choice requires a credible comparator and design, since the unchosen outcome is unobserved.

\subsection{What the data measure}

Each data source reveals some aspects of a player's or team's state and leaves others uncertain. Table~\ref{tab:data} summarizes these measurement limits.

\begin{table}[htbp]
\centering
\caption{Major NBA data layers, their decision-relevant contribution, and their principal boundary.}
\label{tab:data}
\small
\begin{tabularx}{\textwidth}{P{2.6cm}Y Y}
\toprule
\textbf{Data layer} & \textbf{What it makes more observable} & \textbf{What remains unresolved} \\
\midrule
Box score and game logs & Outcomes, usage, basic efficiency, participation, career trajectories & Teammate and opponent context, off-ball contribution, role, and causal attribution \\
Play-by-play and possessions & On/off context, events, substitutions, score and clock states & Within-possession movement, defensive assignment, strategy selection, and hidden coaching information \\
Lineups and rotations & Realized combinations, stint outcomes, substitution patterns & Untried combinations, why lineups were chosen, fatigue, feasibility of alternative rotations \\
Shot and tracking data & Spatial behavior, contest, movement, action sequences, role features & Proprietary access, cross-vendor definitions, tactical intent, transfer to a new team or role \\
Injury and participation records & Reported status, games missed, return dates, schedule and travel exposure & Clinical state, workload dose, reporting changes, medical intervention and selection \\
Prospect and draft data & Pre-draft performance, measurements, market beliefs, selection and career outcomes & Team boards, interviews, workout selection, opportunity after selection, class-specific priors \\
Contract, transaction and asset data & Salary, years, options, reported terms, completed moves, pick ownership & Complete clauses, rejected offers, counterpart preferences, precise historical legal action sets \\
\bottomrule
\end{tabularx}
\end{table}

Tracking data reveal off-ball movement, although tactical intent and portability remain difficult to infer. League medical surveillance can improve injury ascertainment, but public injury reports cannot reproduce it. Transaction records reveal agreed prices while omitting failed offers and opportunities that never reached negotiation. These examples make the observation process part of the decision problem: what was measured, for whom, and when it became available all affect the claim the data can support. Section~\ref{sec:reporting-contract} returns to these questions in the reporting requirements.

\section{Evidence Across Seven Research Streams}

\subsection{On-court action and possession value}

\paragraph{Decision question.} Given the ball, score, clock, player locations, matchup, and opponent response, which action should a player or lineup take?

Possession-based accounting made it possible to separate pace from efficiency and compare events on a common basis \citep{Kubatko2007}. Shot-selection models built on this foundation by weighing shot quality against the opportunity cost of continuing a possession \citep{Skinner2012}. Spatial and tracking studies brought shooting locations, defensive regions, movement, and individual actions into the analysis \citep{Miller2014,FranksDefense2015,Kovalchik2023}. Multiresolution possession and transition models estimate how expected value changes as play unfolds \citep{Cervone2016,Sandholtz2020}. Jutamulia and Hosoi restrict the comparison to immediate pass and shot alternatives. Their approach compares realized actions with modeled opportunities and helps distinguish the creation of opportunities from their execution \citep{Jutamulia2025}.

These studies connect observations to choices at the timescale of play, bringing them close to an action-value framework. Their estimates nevertheless depend on the actions that players and coaches selected and on how the defense responded. Uncommon actions may have little support in the data. A high modeled value gives no assurance that a play can be taught, executed, or sustained against an adapting opponent. Held-out possessions can test predictive performance, but evaluating a coaching intervention requires an identification strategy, credible off-policy evaluation, or prospective implementation. For propensity-weighted evaluation, action probabilities must be logged or defensibly estimated and overlap must be adequate. Rare actions increase variance; unobserved coaching information can undermine the assignment model \citep{SwaminathanJoachims2015}.

Personnel and lineup composition offer a useful next step. A pass may have high expected value because a particular receiver or screener is on the floor, or because the spacing permits it. Estimating action values conditional on those capabilities and on defensive response would make the results more informative for coaches. It would also clarify which recommendations are likely to survive changes in personnel or opponent strategy.

\subsection{Player value across teams and roles}

\paragraph{Decision question.} How much would a player contribute over a future horizon for a particular team, role, and set of teammates?

Adjusted plus-minus methods confront the difficulty of attributing jointly produced outcomes to individual players. Substitutions overlap, minutes are selected, and some players rarely appear apart. Regularization reduces instability from collinearity and limited lineup overlap, while Bayesian models express uncertainty in contributions and rankings \citep{FearnheadTaylor2011,Barrientos2023}. In held-out games, \citet{Sill2010} reduced game-margin RMSE from 12.01 to 11.47 points with regularization, conditional on the games' realized lineups and possessions. This is predictive evidence, but not a forecast using only pregame information. By contrast, \citet{DeshpandeJensen2016} estimate retrospective, context-dependent contributions to winning; they explicitly caution against interpreting those estimates as future player performance. Meta-analytic work further shows that plausible metrics can differ in discrimination, stability, and independence \citep{FranksMeta2016}. Labor-economics research questions whether regularized adjusted plus-minus isolates individual productivity when teammates' contributions are complementary \citep{Ghimire2020}.

The resulting estimates describe contribution after adjustment for the context represented in the model. Using them for recruitment requires an additional inference: that the contribution will persist with different teammates, coaching, or responsibilities. Shrinkage addresses sampling variability, but it cannot establish that inference. Validation needs to reflect the target, whether that is retrospective contribution, near-term performance in the same setting, portable skill, or contribution under a specified team change.

Team and role changes provide opportunities to test transport. A model can be fitted using only information available before a switch, then assessed against subsequent minutes, role, lineup context, and contribution. The assumptions linking the two settings need to be stated explicitly \citep{PearlBareinboim2014}. Calibration and uncertainty coverage matter alongside ranking accuracy: a team needs to know both the contribution projected for its intended deployment and how uncertain that projection is.

\subsection{Roles, capabilities, and assignments}

\paragraph{Decision question.} Which functions can a player perform, which role will be assigned, and how will that role change in a new system?

Traditional positions group players with quite different responsibilities. Tracking features, spatial measures, networks, and clustering describe shooting, creation, screening, passing, rebounding, rim protection, and defensive movement in more detail \citep{SkinnerGuy2015,Muniz2022}. Such representations help compare players, identify hybrid roles, describe roster composition, and provide inputs to performance and acquisition models. The tracking-oriented model of \citet{SkinnerGuy2015}, for example, was demonstrated on 780 hand-coded sequences from three 2011 playoff games. Its limited cross-lineup checks provide evidence about skill transfer within that setting, rather than broad validation after team or coaching changes.

Interpreting those representations requires care. An observed role partly reflects what an organization asks a player to do. A cluster summarizes similarities in measured features, so membership alone does not establish that two players are interchangeable. Roles also change within games and across seasons, teammates, and coaches. A fixed label can obscure the adaptability a team hopes to acquire.

It is useful to distinguish \emph{capability}, the actions a player can perform at an acceptable level; \emph{assignment}, the functions the team asks the player to undertake; and \emph{execution}, what the player does in a particular setting. Longitudinal and multi-view representations could help estimate transitions among these states. Tests on unseen teams and roles would then provide evidence about whether the representation supports the intended use. For roster construction, this means estimating degrees of substitutability among capability bundles, beyond assigning each player a single archetype.

\subsection{Synergy, lineups, and rotation choice}

\paragraph{Decision question.} Which combinations of players create value beyond the sum of individual contributions, and what rotation should be used against a given opponent?

Research on synergy examines interaction effects, productivity spillovers, complementary skills, networks, and lineup performance \citep{Maymin2013,Arcidiacono2017,Kuehn2017,DevlinUminsky2020}. Structural production and possession models also compare hypothetical teammate configurations and trades. These comparisons are substantive counterfactual exercises, but their interpretation depends on assumptions about production and selection; they do not identify the effect of deploying a new lineup without those assumptions. Predictive work uses player features and related representations to estimate the performance of units with little or no shared playing time \citep{Martonosi2023}. Strategic-network studies offer another perspective by describing team process and the concentration of ball movement \citep{Fewell2012}.

Lineup selection complicates all of these approaches. Coaches use information about practice, playbook knowledge, health, matchups, development, and interpersonal considerations that researchers may not observe. A lineup absent from the data could be infeasible, unsuitable, or simply unnecessary in the games played. Even units that appear regularly do so against selected opponents and in selected score states. These features limit what can be learned by comparing their observed outcomes.

The distinction between observed interaction, predicted lineup performance, and interventional lineup value helps specify the problem. \emph{Observed interaction} describes departures from an additive benchmark among players who shared the court. \emph{Predicted lineup performance} concerns a specified unit, including one that has not played. \emph{Interventional lineup value} concerns the change caused by deploying one feasible lineup rather than another. Isolating synergy additionally requires an explicit benchmark for additive individual contributions. Temporal holdouts can assess prediction; intervention effects also require support checks and an identification strategy addressing coaching selection and opponent context. The rotation also matters. A strong five-player unit may be a poor choice if it leaves the team unable to cover the remaining minutes.

\subsection{Availability and workload decisions}

\paragraph{Decision question.} Should a player play, rest, reduce workload, or follow a return-to-play progression, and what is the roster consequence of each choice?

NBA injury studies cover long-run epidemiology, schedule associations, forecasting, and workload measurement \citep{Drakos2010,Mack2019,Teramoto2017,Lewis2018,Cohan2021,Russell2021}. Public analyses commonly draw on missed games, reports, transactions, news, and box-score measures of load. The league's player injury and illness database offers more structured ascertainment, although the data are unavailable for public replication \citep{Mack2019}. Studies of schedules, travel, and time zones add information about environmental exposures \citep{Charest2021}.

These sources measure different aspects of availability. An official designation, a missed game, a minutes restriction, and a clinical diagnosis are not interchangeable. A player may be fit to participate but rested, cleared for limited minutes, or classified under a reporting rule that changed during the sample. Public labels thus reflect both health and organizational reporting decisions. Any model intended to guide participation needs to account for that observation process.

The intervention question is harder. An association between recent load and injury does not establish whether reducing minutes would improve health or how replacement minutes would affect team performance and risk. Research on rest and playoff performance, along with emerging heterogeneous-treatment approaches, begins to address this question \citep{Belk2017,NakamuraSakai2024}. Using internal surveillance data, \citet{Herzog2026} found no statistically significant differences in subsequent regular-season injury risk across rest/load-management groups; the outcome was an in-game injury causing at least two consecutive missed games. The estimates concern that outcome and study population. Rest remained nonrandom, so the result does not establish that an individual rest intervention is ineffective. Participation decisions call for a defined intervention and attention to time-varying confounding, competing risks, recurrent events, and individual uncertainty. For team decisions, the useful output is a joint distribution over participation, workload capacity, performance, recurrence, and replacement quality.

The same issues carry into roster construction. For a team with concentrated salary commitments and limited replacement tools, correlated absences can create substantial downside. Linking availability forecasts to lineup and contract models would allow a team to compare that exposure with the costs and benefits of greater depth. The reviewed literature offers limited validation of this connection.

\subsection{Drafting and development opportunities}

\paragraph{Decision question.} Which prospect should a team select or acquire, at what pick price, and how should that player be developed after selection?

Draft studies use college production, measurements, combine results, and mock drafts to predict selection, early performance, career outcomes, and market consensus \citep{Berri2011,BergerDaumann2021,FisherMontague2025,IchniowskiPreston2017}. These targets capture different processes. Draft order reflects team evaluations and broader market expectations; early minutes depend partly on the opportunity a team provides; and career contribution combines ability with health, coaching, development, and survival in the league. A model can predict draft position accurately while reproducing biases embedded in historical team selection decisions. The studies also differ in validation: \citet{Berri2011} report retrospective regressions without held-out predictive evaluation, whereas \citet{FisherMontague2025} evaluate dated mock drafts against actual selection order.

Associations between draft order, later playing time, and career survival illustrate the difficulty of separating prospect quality from organizational opportunity \citep{StawHoang1995}. If realized career production is the sole outcome, a model may receive credit for anticipating the selection process rather than improving on it. Draft evaluation should trace pre-draft signals through market beliefs, the team's choice, development opportunities, and subsequent contribution. Missing combine tests, workouts, and interviews also require interpretation as selection states, rather than being entered as zeroes.

The pick itself changes the decision. Selecting a prospect fourth and selecting that prospect twenty-fourth involve different opportunity costs, contracts, alternative boards, and trade possibilities. A useful evaluation would preserve the information available at the pick and compare player outcomes together with asset cost. Archived consensus boards and validation across draft classes are especially valuable because both class strength and the information available to teams change over time. The retained literature is stronger on prospect selection and post-selection opportunity than on evaluations of development interventions; the stream label should not imply equal coverage of both.

\subsection{Contracts, acquisition, and roster construction}

\paragraph{Decision question.} Should a team sign, extend, trade, waive, or retain a player, and which legally and commercially feasible portfolio best serves its objectives?

Salary research examines compensation, discrimination, offensive and defensive returns, and pre-contract performance \citep{Ehrlich2019,Wen2023,Kaplan2024}. Return-on-investment measures combine performance and salary to estimate apparent surplus \citep{Lautier2025}. These results describe how the labor market rewards players, but a team deciding whether to offer a contract faces a different question: what would the player be worth in its roster? Bargaining, maximum and minimum salary rules, service, timing, cap conditions, and scarce alternatives all influence the observed price. Salary is therefore an outcome to be explained as well as a cost to be paid.

Acquisition and roster models approach this choice more directly, combining player metrics with role balance, network fit, or optimization \citep{Brill2023,MunizFlamand2023,Ke2024,MayminGM2017}. Work on risk and uncertainty places team building in a dynamic setting \citep{SchmidtTeamBuilding2021}; a recent preprint develops a rolling-horizon stochastic formulation \citep{Zhang2026}. The practical difficulty is defining the actions a team can take. A payroll constraint captures only part of the CBA. Transactions also depend on roster slots, contract types, rights, exceptions, salary matching, hard caps, aprons, draft assets, timing, and counterpart willingness. Bringing these conditions into the candidate generator would allow optimization to address a more realistic set of choices.

Legal permission and market availability remain separate even in a detailed rule model. A permitted trade may have no willing counterparty. Observed trades are selected agreements, while a non-trade could represent rejection, lack of availability, a dominated option, or a proposal never considered. Roster models need some account of these possibilities, whether through probabilities, bounds, or clearly labeled assumptions. They also need to carry uncertainty in player states and the value of future options into the comparison.

\subsection{Cross-stream synthesis}

Across the seven streams, research tends to provide stronger support when the target stays close to the observation unit: a possession, stint, game, draft selection, or contract. More assumptions are needed as an estimate is transferred to a new role, used to choose a workload, or combined with other estimates in an acquisition decision. Selection and feasibility then become prominent. Coaches, medical staff, teams, prospects, and counterparties influence which actions appear in the data and which alternatives were available. The following section examines these shared difficulties.

\section{Connecting Estimates to Decisions:\texorpdfstring{\\}{ }Strong States, Weak Interfaces}

Figure~\ref{fig:relationships} maps the relationships between measurement, player states, and decisions. Tracking and spatial-defense data inform contribution and synergy as well as role. Possession models contribute to player evaluation and tactical choice; lineup and passing networks describe roles and complementarity; and schedule and participation records inform availability \citep{FranksDefense2015,SkinnerGuy2015,Cervone2016,Fewell2012,Muniz2022,Mack2019}. The less established links involve combining these estimates, applying them to a different setting, and comparing actions subject to the rules and opportunities at the time.

\begin{figure}[htbp]
  \centering
  \resizebox{0.98\textwidth}{!}{\begin{tikzpicture}[
  every node/.append style={execute at begin node={\hyphenpenalty=10000\relax\exhyphenpenalty=10000\relax}},
  font=\footnotesize,
  panel/.style={rounded corners=4pt, draw=grayline, fill=grayfill, line width=0.7pt},
  grid/.style={draw=grayline, line width=0.45pt},
  primary/.style={circle, draw=navy, fill=navy, minimum size=6pt, inner sep=0pt},
  support/.style={circle, draw=teal, fill=tealfill, line width=0.8pt,
                  minimum size=6pt, inner sep=0pt},
  title/.style={font=\small\bfseries, text=navy},
  rowlabel/.style={anchor=west, text width=4.4cm, align=left},
  head/.style={font=\scriptsize\bfseries, align=center, text=graytext}
]
  \node[title] at (0,5.45) {Selected measurements $\longrightarrow$ decision-relevant states};
  \node[panel, minimum width=15cm, minimum height=4.95cm] at (0,2.475) {};
  \draw[grid] (-7.5,4.95) rectangle (7.5,0);
  \draw[grid] (-2.6,4.95) -- (-2.6,0);
  \foreach \x in {-0.65,1.30,3.25,5.20} {\draw[grid] (\x,4.95)--(\x,0);}
  \foreach \y in {4.00,3.20,2.40,1.60,0.80} {\draw[grid] (-7.5,\y)--(7.5,\y);}

  \node[head] at (-1.625,4.475) {player\\contribution};
  \node[head] at (0.325,4.475) {role /\\capability};
  \node[head] at (2.275,4.475) {synergy /\\fit};
  \node[head] at (4.225,4.475) {availability /\\capacity};
  \node[head] at (6.350,4.475) {market value /\\development};

  \node[rowlabel] at (-7.25,3.60) {box score and plus-minus};
  \node[rowlabel] at (-7.25,2.80) {tracking, spacing and defense};
  \node[rowlabel] at (-7.25,2.00) {possession and action value};
  \node[rowlabel] at (-7.25,1.20) {lineup and passing networks};
  \node[rowlabel] at (-7.25,0.40) {schedule, injury and\\participation records};

  \node[primary] at (-1.625,3.60) {}; \node[support] at (0.325,3.60) {};
  \node[support] at (2.275,3.60) {}; \node[support] at (6.350,3.60) {};
  \node[primary] at (-1.625,2.80) {}; \node[primary] at (0.325,2.80) {};
  \node[primary] at (2.275,2.80) {}; \node[support] at (6.350,2.80) {};
  \node[primary] at (-1.625,2.00) {}; \node[primary] at (0.325,2.00) {};
  \node[support] at (2.275,2.00) {};
  \node[support] at (-1.625,1.20) {}; \node[primary] at (0.325,1.20) {};
  \node[primary] at (2.275,1.20) {}; \node[support] at (6.350,1.20) {};
  \node[support] at (-1.625,0.40) {}; \node[primary] at (4.225,0.40) {};
  \node[support] at (6.350,0.40) {};

  \node[title] at (0,-0.65) {Decision-relevant states $\longrightarrow$ basketball-operations decisions};
  \node[panel, minimum width=15cm, minimum height=4.95cm] at (0,-3.575) {};
  \draw[grid] (-7.5,-1.10) rectangle (7.5,-6.05);
  \draw[grid] (-2.6,-1.10) -- (-2.6,-6.05);
  \foreach \x in {-0.65,1.30,3.25,5.20} {\draw[grid] (\x,-1.10)--(\x,-6.05);}
  \foreach \y in {-2.05,-2.85,-3.65,-4.45,-5.25} {\draw[grid] (-7.5,\y)--(7.5,\y);}

  \node[head] at (-1.625,-1.575) {lineup /\\tactics};
  \node[head] at (0.325,-1.575) {rest /\\workload};
  \node[head] at (2.275,-1.575) {draft /\\development};
  \node[head] at (4.225,-1.575) {contract /\\trade};
  \node[head] at (6.350,-1.575) {dynamic roster\\construction};

  \node[rowlabel] at (-7.25,-2.45) {player contribution};
  \node[rowlabel] at (-7.25,-3.25) {role and capability};
  \node[rowlabel] at (-7.25,-4.05) {synergy and lineup fit};
  \node[rowlabel] at (-7.25,-4.85) {availability and capacity};
  \node[rowlabel] at (-7.25,-5.65) {market value and development};

  \node[primary] at (-1.625,-2.45) {}; \node[support] at (0.325,-2.45) {};
  \node[primary] at (2.275,-2.45) {}; \node[primary] at (4.225,-2.45) {};
  \node[primary] at (6.350,-2.45) {};
  \node[primary] at (-1.625,-3.25) {}; \node[support] at (0.325,-3.25) {};
  \node[primary] at (2.275,-3.25) {}; \node[primary] at (4.225,-3.25) {};
  \node[primary] at (6.350,-3.25) {};
  \node[primary] at (-1.625,-4.05) {}; \node[support] at (2.275,-4.05) {};
  \node[support] at (4.225,-4.05) {}; \node[primary] at (6.350,-4.05) {};
  \node[primary] at (-1.625,-4.85) {}; \node[primary] at (0.325,-4.85) {};
  \node[support] at (2.275,-4.85) {}; \node[primary] at (4.225,-4.85) {};
  \node[primary] at (6.350,-4.85) {};
  \node[support] at (-1.625,-5.65) {}; \node[primary] at (2.275,-5.65) {};
  \node[primary] at (4.225,-5.65) {}; \node[primary] at (6.350,-5.65) {};

  \node[primary] at (-6.25,-6.72) {};
  \node[anchor=west] at (-6.05,-6.72) {primary or direct use};
  \node[support] at (0.95,-6.72) {};
  \node[anchor=west] at (1.15,-6.72) {supporting or contextual use};
  \node[align=center, text=graytext, font=\scriptsize] at (0,-7.18)
    {Blank cells indicate no standard direct interface in the reviewed literature, not a claim of impossibility.};
\end{tikzpicture}}
  \caption{Relationships among major NBA analytics research streams. The upper matrix records how selected measurement streams inform decision-relevant player states; the lower matrix records how those states enter basketball-operations decisions. Filled markers denote primary or direct uses in representative studies, outlined markers denote supporting or contextual uses, and blank cells indicate that the review did not identify a standard direct interface. The map is conceptual rather than a citation network, maturity score, or paper count.}
  \label{fig:relationships}
\end{figure}

\subsection{Aligning observation and decision units}

The observation unit often differs from the decision unit. Studies analyze player-seasons, stints, possessions, games, or contracts; a team chooses an action concerning a player in a particular setting and at a particular date. Bridging the two takes additional work. A player-season prediction needs a deployment assumption, a lineup-stint prediction needs a rotation, and a salary prediction needs the alternatives available in negotiation. Specifying the decision owner, timestamp, candidate actions, and outcome horizon helps identify which links the analysis must supply.

The interpretation of the target matters just as much. Observed contribution, future performance, treatment response, and the value of information are different estimands. A recommendation cannot be justified by moving between them implicitly. Where causal identification is unavailable, scenario analysis or partial identification can establish the conditions under which an action would be preferred.

\begin{samepage}
\subsection{Reconstructing the opportunity set}

Researchers generally observe selected actions: the lineup a coach used, the prospect a team drafted, the player rested, or the contract both sides signed. The alternatives are much less visible. Coding every unrealized action as a negative observation obscures the difference between an option that was rejected and one that was never available.

\end{samepage}

Some alternatives can be reconstructed from dated rules and public information. Selection can also be studied using observed decision variables, instruments, quasi-experiments, or explicit structural assumptions. When the opportunity set remains uncertain, partial identification and sensitivity bounds can show how much the conclusion depends on it \citep{Manski2003}. The choice of method should reflect what is known about availability, rather than assume that every player-team pair or hypothetical trade was possible.

\subsection{Information available at the decision date}

Information changes between an event and the date a researcher retrieves the data. Injury designations may be updated several times before tip-off; contract terms may become public after an agreement; and a consensus draft board may incorporate news released after a selection. Replaying a decision with a later file can give the model information the team did not have. Retrospective analyses therefore need versioned snapshots that distinguish event, release, and retrieval times from the model cutoff. Rolling-origin evaluation is appropriate for this temporal forecasting task; random folds spanning seasons may allow later information into earlier predictions \citep{Tashman2000}.

Changes in tracking definitions, reporting practices, participation policies, the CBA, and lottery rules can also alter what the data mean. A model spanning several regimes needs to represent those differences and assess how well its estimates transfer across them. Treating the sample as stationary would leave this part of the claim untested.

Table~\ref{tab:readiness} summarizes the six decision-readiness gates and their diagnostic questions.

\begin{table}[!htbp]
\centering
\caption{Six decision-readiness gates and their diagnostic questions.}
\label{tab:readiness}
\small
\begin{tabularx}{\textwidth}{P{2.7cm}Y Y}
\toprule
\textbf{Gate} & \textbf{Diagnostic question} & \textbf{Common failure} \\
\midrule
Point-in-time validity & Were inputs and definitions available at the decision time? & Later injury status, corrected data, or post-draft information leaks into training or replay \\
Uncertainty & Does the decision rule use calibrated distributions, intervals, or scenarios? & A recommendation is based on a noisy point estimate without propagating decision-relevant uncertainty \\
Context portability & Is transfer to the intended team, role, lineup, opponent, or regime tested? & Observed contribution is treated as intrinsic and invariant ability \\
Action feasibility & Are actions legal and operationally possible at that date? & An optimizer uses only a payroll cap or ignores the full rotation and contract rights \\
Opportunity-set observability & Are available alternatives observed, reconstructed, or bounded? & Untried lineups, undrafted players, and non-trades are treated as ordinary negatives \\
Evaluation & Is the selected action evaluated against credible alternatives under conditions appropriate to the decision claim? & Predictive accuracy or a realized retrospective outcome is treated as evidence that the recommended action improved the decision \\
\bottomrule
\end{tabularx}
\vspace{0.35em}

\parbox{\textwidth}{\footnotesize\textit{Accountability note.} Public reproducibility and organizational auditability are separate reporting requirements rather than a seventh decision-readiness gate: they determine whether the evidence and handoffs can be inspected, not whether an action improved a decision.}
\end{table}

\subsection{Carrying uncertainty into the choice}

Uncertainty can materially alter an otherwise attractive choice. A low-probability absence may be costly for a shallow roster; a risk of role failure may matter over a long contract; and correlated injuries can undermine apparent depth. Draft-pick protections introduce further asymmetry. Reporting uncertainty at the estimation stage is insufficient if the recommendation then treats the forecast as certain.

The comparison of actions can use posterior distributions, calibrated probabilities, intervals, or scenarios, depending on the uncertainty at issue. Conformal methods, scenario trees, robust optimization \citep{BertsimasBrownCaramanis2011}, and value-of-information analysis \citep{Howard1966} offer different ways to examine it. The useful test is whether the analysis identifies conditions that change the preferred action or establishes its robustness across plausible conditions. Elaborate uncertainty modeling contributes little when the candidate set is unrealistic, whereas simple scenarios may reveal a consequential weakness.

\subsection{Research with public and restricted data}

Much of the information used inside teams is private: raw tracking, medical surveillance, workload sensors, scouting, practice records, team boards, contract clauses, negotiations, and rejected offers. Public research can still test benchmarks and biases, compare reduced public-data models, and release definitions and code. For a decision that depends on inaccessible information, researchers can reconstruct the legal and publicly visible options, specify plausible ranges for missing health or market states, and report how the preferred action changes across those ranges. The claim should be limited to the region supported by the analysis.

For internal research, provenance, versioning, and independent audit provide ways to scrutinize findings that cannot be publicly replicated. Public reproducibility and internal auditability answer different questions about access and accountability. Neither, on its own, establishes that the model supports the proposed decision.

Figure~\ref{fig:gapmap} summarizes the maturity of these interfaces in the retained literature. Measurement interfaces are developed in several streams. Predictive modeling is present across all seven streams, but prediction is rated Developed only for on-court action and Emerging elsewhere, reflecting limited accumulation or validation for the specified targets. Causal and model-based counterfactual work is developing unevenly and retains substantial identification limits. Constrained prescription has a smaller evidence base, and no retained study prospectively evaluates an implemented NBA organizational decision rule against a credible alternative. These findings concern the reviewed evidence; they cannot establish the extent of private team practice or of research outside the corpus.

\begin{figure}[H]
  \centering
  \resizebox{0.95\textwidth}{!}{\begin{tikzpicture}[font=\scriptsize]
  \def\cw{2.15}
  \def\rh{0.78}
  \def\xzero{4.0}
  \def\ytop{4.8}
  \newcommand{\gapcell}[4]{%
    \path[draw=white, line width=1.5pt, fill=#3]
      (\xzero+#1*\cw,\ytop-#2*\rh-\rh) rectangle ++(\cw,\rh);
    \node[text=graytext] at (\xzero+#1*\cw+0.5*\cw,\ytop-#2*\rh-0.5*\rh) {#4};
  }

  \node[align=center, font=\scriptsize\bfseries, text width=2.08cm] at (\xzero+0.5*\cw,\ytop+0.55) {Measurement};
  \node[align=center, font=\scriptsize\bfseries, text width=1.9cm] at (\xzero+1.5*\cw,\ytop+0.55) {Prediction};
  \node[align=center, font=\scriptsize\bfseries, text width=2.18cm] at (\xzero+2.5*\cw,\ytop+0.55) {Causal /\\counterfactual};
  \node[align=center, font=\scriptsize\bfseries, text width=2.1cm] at (\xzero+3.5*\cw,\ytop+0.55) {Constrained\\prescription};
  \node[align=center, font=\scriptsize\bfseries, text width=1.9cm] at (\xzero+4.5*\cw,\ytop+0.55) {Prospective\\evaluation};

  \node[anchor=east, font=\scriptsize\bfseries] at (\xzero-0.18,\ytop-0.5*\rh) {On-court action};
  \node[anchor=east, font=\scriptsize\bfseries] at (\xzero-0.18,\ytop-1.5*\rh) {Player value};
  \node[anchor=east, font=\scriptsize\bfseries] at (\xzero-0.18,\ytop-2.5*\rh) {Role / capability};
  \node[anchor=east, font=\scriptsize\bfseries] at (\xzero-0.18,\ytop-3.5*\rh) {Synergy / lineup};
  \node[anchor=east, font=\scriptsize\bfseries] at (\xzero-0.18,\ytop-4.5*\rh) {Availability};
  \node[anchor=east, font=\scriptsize\bfseries] at (\xzero-0.18,\ytop-5.5*\rh) {Draft / development};
  \node[anchor=east, font=\scriptsize\bfseries] at (\xzero-0.18,\ytop-6.5*\rh) {Contract / roster};

  \gapcell{0}{0}{strongfill}{developed}\gapcell{1}{0}{strongfill}{developed}
  \gapcell{2}{0}{emergingfill}{emerging}\gapcell{3}{0}{emergingfill}{emerging}\gapcell{4}{0}{thinfill}{thin}
  \gapcell{0}{1}{strongfill}{developed}\gapcell{1}{1}{emergingfill}{emerging}
  \gapcell{2}{1}{emergingfill}{emerging}\gapcell{3}{1}{thinfill}{thin}\gapcell{4}{1}{thinfill}{thin}
  \gapcell{0}{2}{strongfill}{developed}\gapcell{1}{2}{emergingfill}{emerging}
  \gapcell{2}{2}{thinfill}{thin}\gapcell{3}{2}{thinfill}{thin}\gapcell{4}{2}{thinfill}{thin}
  \gapcell{0}{3}{strongfill}{developed}\gapcell{1}{3}{emergingfill}{emerging}
  \gapcell{2}{3}{emergingfill}{emerging}\gapcell{3}{3}{emergingfill}{emerging}\gapcell{4}{3}{thinfill}{thin}
  \gapcell{0}{4}{emergingfill}{emerging}\gapcell{1}{4}{emergingfill}{emerging}
  \gapcell{2}{4}{emergingfill}{emerging}\gapcell{3}{4}{thinfill}{thin}\gapcell{4}{4}{thinfill}{thin}
  \gapcell{0}{5}{emergingfill}{emerging}\gapcell{1}{5}{emergingfill}{emerging}
  \gapcell{2}{5}{emergingfill}{emerging}\gapcell{3}{5}{thinfill}{thin}\gapcell{4}{5}{thinfill}{thin}
  \gapcell{0}{6}{emergingfill}{emerging}\gapcell{1}{6}{emergingfill}{emerging}
  \gapcell{2}{6}{emergingfill}{emerging}\gapcell{3}{6}{emergingfill}{emerging}\gapcell{4}{6}{thinfill}{thin}

  \draw[draw=grayline, line width=0.6pt]
    (\xzero,\ytop-7*\rh) rectangle (\xzero+5*\cw,\ytop);

  \node[font=\scriptsize, text width=11.0cm, align=center] at (\xzero+2.5*\cw,-1.75)
    {\textbf{Interpretation:} maturity of specified interfaces in the retained public literature. Model-based counterfactuals need not identify causal effects. The recurring gap is the move to dated, feasible and prospectively evaluated action.};
\end{tikzpicture}}
  \caption{Qualitative evidence-gap map. Labels characterize each specified interface within the retained public literature, not the domain as a whole, study quality, or publication volume. Developed denotes an established research line with validation appropriate to its claim level; emerging denotes substantive direct work with limited accumulation or validation; thin denotes mainly isolated, indirect, or conceptual evidence. The causal/counterfactual column includes model-based comparisons whose identification assumptions remain unverified. Coding rules and cell-level targets, evidence, and boundaries appear in Supplementary Section~S3 and Table~S3; report-level mappings appear in Table~S4.}
  \label{fig:gapmap}
\end{figure}

\FloatBarrier
\section{Research Opportunities Under Changing NBA Rules}

Applying the framework to recent CBA and Draft Lottery changes shows how it generates research questions from a changing decision environment. Both affect the rule state $R_t$, the feasible set $\mathcal{A}$, and the incentives represented in Equation~\ref{eq:decision}. They also change how historical data should be interpreted. For effects that have yet to occur, researchers have an opportunity to preserve data and specify hypotheses before observing outcomes.

\subsection{The 2023 CBA and the value of transaction options}

The CBA took effect on 1 July 2023 and runs through the 2029--30 season, with either side able to opt out after 2028--29 \citep{NBA2023CBA,NBA2024CBA101}. Its apron rules restrict transaction tools in specified team states, including salary aggregation, exceptions, and the use of future draft assets. The Frozen Pick rule illustrates the longer-term consequences. Beginning in 2024--25, Second-Apron status at the start of a team's final regular-season game can freeze its first-round pick in the seventh Draft after the end of that Salary Cap Year. Exposure at that measurement point in at least two of the next four cap years can move the pick to the end of the first round and keep it untradeable \citep{NBA2024CBA101}. An acquisition can therefore change a team's future options as well as its payroll and playing strength. Its value depends on the restrictions the team would face.

Four related lines of research follow from these constraints.

\begin{enumerate}
  \item \textbf{Team-specific player decision value.} Replace league-wide player-surplus rankings with a player-by-team-by-date distribution that combines contribution, role fit, joint availability, contract cash flows, rights, and the shadow price of lost transaction flexibility.
  \item \textbf{Role-bundle substitution.} Test whether combinations of lower-cost capabilities can replace a scarce, expensive function under apron constraints. The target is not similarity between player archetypes but the ability of a feasible bundle to reproduce lineup functions over a full rotation.
  \item \textbf{Availability-adjusted roster fragility.} Compare star concentration and depth strategies using correlated participation and workload-capacity scenarios, replacement quality, legal midseason tools, and downside-risk measures. Mean projected wins are insufficient when the distribution is asymmetric.
  \item \textbf{The causal value of transaction rights.} Reconstruct exact team-date apron and contract states and estimate how losing particular transaction tools changes signings, trades, roster continuity, depth, and performance. Payroll above a threshold is an exposure, not the causal mechanism; the target is the restriction or option whose loss changes behavior.
\end{enumerate}

These programs overlap with the broader priorities in Table~\ref{tab:agenda}, since the same apron restriction can affect player transport, substitution, roster risk, acquisition options, and rights valuation. A versioned representation of the CBA would make those connections easier to study. Causal evaluation remains difficult, however. Teams may adjust contracts before a restriction binds, and exposure varies with team quality, ownership spending, and competitive timeline. Difference-in-differences designs need treatment-timing diagnostics; matched team-date comparisons and event studies need to address anticipation. Structural or simulation models offer another route, provided they make the assumptions used to value flexibility explicit.

\subsection{Prospective research on the 3-2-1 Lottery}

On 28 May 2026, the NBA approved a 3-2-1 Draft Lottery for the 2027, 2028, and 2029 drafts \citep{NBA2026Lottery}, with the stated aim of reducing incentives to prioritize draft position over winning. The lottery expands to 16 teams. Non-playoff and non-Play-In teams generally receive three balls, except the bottom three teams, which receive two. The No. 9 and No. 10 Play-In seeds in each conference receive two, and the losers of the No. 7-versus-No. 8 Play-In games receive one. The drawing orders all first 16 picks. The bottom three teams cannot fall below pick 12, while the other lottery teams can fall as far as pick 16. Additional provisions limit repeated high own-pick outcomes, restrict certain new pick protections, and expand disciplinary authority. These are changes to the rules. Their effects on the value of moving up or down the standings remain questions for empirical research.

No regular season or draft had been conducted under the new system by the evidence cutoff of 1 September 2026. Game-level analyses can therefore be planned before the first affected season. Historical work documents responses to draft incentives and the difficulty of attributing behavior to lottery changes \citep{TaylorTrogdon2002,Price2010,SchmidtLottery2024}. For the new system, a binary ``tanking'' measure would discard much of the relevant variation. One possible treatment is the rule-induced change in the expected value of a team's own or economically exposed draft asset from an additional loss at a given date. Calculating it requires standings, schedules, ownership, swaps, protections, and an estimate of prospect-class value, together with the full rank-conditioned pick distribution incorporating pick floors and repeated-pick restrictions.

Availability-adjusted lineup strength, veteran shutdowns, young-player minutes, transactions, and late-game choices are possible outcomes, although none directly reveals intent. A credible design would specify these measures in advance and examine placebo ranking boundaries where the lottery rule does not change, alternative pick-value curves, simulated schedules, injuries, and playoff incentives. The two-ball allocation for the bottom three teams and the inclusion of Play-In teams introduce incentive kinks. Because ranks can be influenced by team behavior and local samples are small, analyses should examine difference-in-discontinuities or simulated, synthetic, and matched comparisons before relying on local randomization.

The relevant exposure date is outcome-specific: asset-market responses may begin at announcement, whereas game-level responses should be examined during the first affected season. Baseline periods should also be checked for anticipation before the announcement. A drawing that orders all 16 picks changes the distribution of pick outcomes, while protection limits and repeated-pick restrictions affect value across years. Trades usually bundle players, picks, swaps, and protections, making the price of any one component difficult to isolate. Interval-valued hedonic models and matched-package comparisons, supported by structural sensitivity analysis, could express that uncertainty more faithfully than a single estimated pick price.

\begin{figure}[htbp]
  \centering
  \resizebox{0.95\textwidth}{!}{\begin{tikzpicture}[
  every node/.append style={execute at begin node={\hyphenpenalty=10000\relax\exhyphenpenalty=10000\relax}},
  font=\small,
  event/.style={rounded corners=3pt, draw=navy, fill=bluefill, text width=2.75cm,
                minimum height=1.25cm, align=center, inner sep=4pt},
  future/.style={rounded corners=3pt, draw=redaccent, fill=redfill, text width=2.75cm,
                 minimum height=1.25cm, align=center, inner sep=4pt},
  line/.style={->, >=latex, draw=graytext, line width=1.2pt},
  date/.style={font=\scriptsize\bfseries, text=navy}
]
  \draw[line] (-7.3,0) -- (7.3,0);
  \foreach \x in {-6,-3,0,3,6} {\draw[graytext, line width=0.8pt] (\x,-0.12)--(\x,0.12);}

  \node[event] (cba) at (-6,1.45) {2023 CBA takes effect\\new apron architecture};
  \node[date] at (-6,-0.38) {July 2023};
  \draw[graytext] (-6,0.12)--(cba.south);

  \node[event] (phase) at (-3,-2.15) {phased transaction and future-pick consequences become observable};
  \node[date] at (-3,0.38) {2024--26};
  \draw[graytext] (-3,-0.12)--(phase.north);

  \node[event] (approve) at (0,1.45) {3-2-1 Lottery approved\\asset-market responses may begin};
  \node[date] at (0,-0.38) {28 May 2026};
  \draw[graytext] (0,0.12)--(approve.south);

  \node[future] (first) at (3,-2.15) {first affected\\playing season\\and 2027 Draft};
  \node[date] at (3,0.38) {2026--27};
  \draw[graytext] (3,-0.12)--(first.north);

  \node[future] (multi) at (6,1.45) {three treated drafts\\first multi-season synthesis};
  \node[date] at (6,-0.38) {through 2029};
  \draw[graytext] (6,0.12)--(multi.south);

  \node[align=center, text width=15cm, font=\scriptsize\bfseries, text=redaccent] at (0,-4.25)
    {Align baseline data with outcome-specific exposure dates; preregister analyses of outcomes not yet observed.};
\end{tikzpicture}}
  \caption{Institutional research timeline. The CBA already produces observable team exposure. The 3-2-1 Lottery applies from the 2027 Draft, with the 2026--27 season the first affected playing season. Asset-market responses may begin at the May 2026 announcement, so baseline periods must be defined separately for each outcome.}
  \label{fig:timeline}
\end{figure}
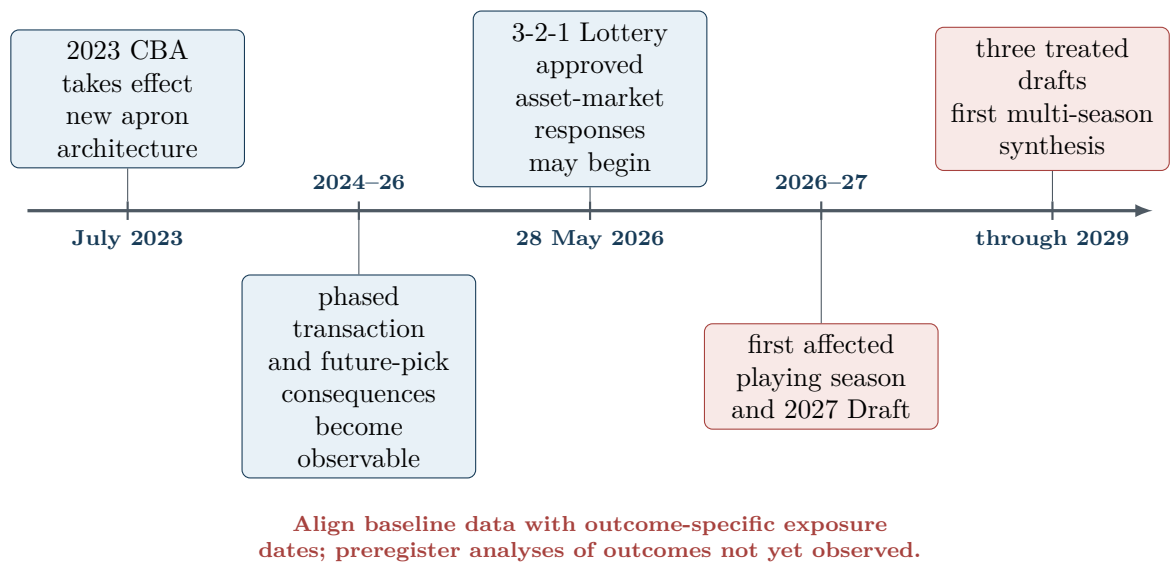

Figure~\ref{fig:timeline} summarizes the timing of these institutional changes and the corresponding research opportunities.

\FloatBarrier
\subsection{Priorities for linked research}

Table~\ref{tab:agenda} brings the proposed studies together. For each, it identifies the estimand, minimum linked data, and an evaluation design, so that the proposal can be assessed against the information it would actually require.

\begin{table}[!htb]
\centering
\caption{Priority research programs generated by the decision-centered synthesis.}
\label{tab:agenda}
\scriptsize
\begin{tabularx}{\textwidth}{P{2.35cm}Y Y Y}
\toprule
\textbf{Program} & \textbf{Decision estimand} & \textbf{Minimum linked data} & \textbf{Credible design or evaluation} \\
\midrule
Player-team transport & Future contribution distribution under a declared team, role, lineup and minutes horizon & Point-in-time player features, role, tracking where available, lineups, availability, team changes & Team-switch and role-switch holdouts; calibration; support checks; uncertainty coverage \\
Role-bundle substitution & Value of a feasible capability bundle relative to the best accessible alternative & Capability features, role assignments, full rotations, salaries, contracts, legal acquisition state & Out-of-team validation plus constrained roster replay; ablation of role and synergy interfaces \\
Roster fragility & Tail loss from correlated unavailability under a roster and replacement policy & Participation, minutes capacity, performance, depth, contracts, CBA replacement tools & Model: joint scenarios and a declared downside objective. Checks: calibration, coverage, stress sensitivity, and replay under stated assumptions \\
Legal and market action generation & Expected value of the best legal and plausibly available personnel action (signing, trade, waiver, or contract extension) & Versioned CBA state, contracts, rights, assets, team needs, transactions and reported offers & Rule-engine tests; candidate-set audit; bounds on market availability; prospectively logged shadow-mode comparisons \\
CBA restriction effects & Effect and shadow value of losing a specified transaction right & Team-date apron state, contract structure, hard-cap triggers, transactions, roster and performance outcomes & Event study or matched panel with anticipation checks; structural sensitivity for option value \\
Continuous lottery incentives & Behavioral response per marginal expected team asset value of a loss & Pregame standings, schedule, roster health, lineup strength, asset exposure, and rule-encoded pick distributions including floors & Pre-registered continuous treatment; alternative pick curves; placebo dates and ranking boundaries without rule changes \\
Pick portfolio valuation & Joint option value of picks, swaps, protections and repeated-pick restrictions & Complete asset graph, dated trade packages, rule state, prospect-class priors and later outcomes & Interval-valued hedonic or structural model; interference and equilibrium sensitivity \\
\bottomrule
\end{tabularx}
\end{table}

In shadow-mode comparisons, candidate sets and recommendations are frozen and logged before outcomes, without requiring the recommended action to be executed. This permits prospective checks of forecasts and action-set records; it does not identify effects of unchosen actions or demonstrate gains from deployment. Several of these studies would supply inputs for others. Estimates that transport across teams could support role-bundle and roster comparisons. Joint availability scenarios could inform contract risk, while a dated legal action generator could help identify exposure to CBA restrictions. An asset graph and rule encoder would serve both lottery-incentive and pick-valuation research. Shared definitions and compatible timestamps would make these connections more useful than a collection of independent model rankings.

\FloatBarrier
\section{Using the Framework in Team Decisions}
\label{sec:team-decisions}

In a team setting, an estimate usually passes through several people before it affects a decision. An analyst's contribution forecast needs to specify the role and minutes assumed. A coach using an availability forecast needs alternatives for workload and replacement minutes. Basketball-operations staff need to assess a proposed roster move against legality, price, market access, risk, and lost flexibility. Table~\ref{tab:practitioners} describes the information needed at these handoffs.

\begin{table}[htbp]
\centering
\caption{Practitioner interfaces for decision-ready NBA analytics.}
\label{tab:practitioners}
\footnotesize
\begin{tabularx}{\textwidth}{P{1.9cm}P{3.0cm}Y Y}
\toprule
\textbf{Primary user} & \textbf{Decision question} & \textbf{Minimum analytical output} & \textbf{Required handoff} \\
\midrule
Coach and performance staff & Who should play, in which role and lineup, at what workload, against this opponent? & Conditional action or lineup values; availability and minutes-capacity scenarios; uncertainty and opponent response & Assumptions that can be translated into a rotation, tactical instruction, or rest plan and observed after implementation \\
General manager and basketball operations & Which feasible acquisition, retention, development, or asset action best advances the team's timeline? & Player--team value distribution; contract and rights state; legal and plausible alternatives; multi-period risk and flexibility & A dated decision memo that separates model value, market price, legal feasibility, and unobserved counterpart willingness \\
Analyst and data engineer & Is the output valid for the decision date, population, and context? & Versioned inputs; target and estimand; calibration; transport and support diagnostics; action-set audit & Reproducible model card, scenario output, data lineage, and a record of what information was actually delivered \\
\bottomrule
\end{tabularx}
\end{table}

\FloatBarrier
\subsection{A proposed deployment workflow}

The framework can be used without building an end-to-end optimizer. The following proposed deployment workflow records the information needed to make and subsequently examine a decision.

\begin{enumerate}
  \item \textbf{Freeze the decision.} Record the owner, timestamp, objective, horizon, and information that was available.
  \item \textbf{Estimate coupled states.} Produce distributions for the components that could change the choice---contribution, role, synergy, availability, cost, or development---and expose their dependencies.
  \item \textbf{Construct candidate actions.} Apply roster, contract, CBA, timing, and basketball constraints; label market availability as observed, assumed, or bounded.
  \item \textbf{Stress-test the ranking.} Vary the assumptions that matter to the user, including minutes, role, opponent, availability, price, and competitive timeline.
  \item \textbf{Issue a decision record.} Report the recommended action, credible alternatives, uncertainty, assumptions, veto conditions, and information that would change the recommendation.
  \item \textbf{Evaluate after the decision.} Preserve the selected action and contemporaneous forecast, then evaluate calibration, process, and outcomes without redefining the target after results are known.
\end{enumerate}

Keeping this record would make it easier to investigate a poor outcome: the cause might lie in the forecast, an unavailable alternative, a failed transport assumption, or an incorrectly encoded rule. The outcome might also have been plausible under a well-calibrated forecast.

\subsection{A reporting contract for decision claims}
\label{sec:reporting-contract}

A reporting contract makes the assumptions behind a decision claim available to its readers and users. Its detail should be proportionate to the claim: a descriptive spatial study needs a clear sample and interpretation, while a recommendation needs evidence about the alternatives and their consequences. For work that supports or implies a decision, we propose the following minimum information.

\begin{enumerate}
  \item \textbf{Decision owner and action.} Identify who acts and what choice is compared: deploy, rest, develop, draft, sign, trade, retain, or allocate an asset.
  \item \textbf{Decision timestamp and information set.} Freeze the latest permissible input, report event and release times, and audit later corrections or labels.
  \item \textbf{Target and estimand.} Distinguish observed contribution, future prediction, causal effect, treatment response, market price, and team-specific decision value.
  \item \textbf{Unit and horizon.} Align the observation unit with the action and outcome horizon. Explain how possession, game, stint, player-season, contract, and team-date units are connected.
  \item \textbf{Uncertainty and calibration.} Report the distributional information that could change a choice, including interval coverage, calibration, scenarios, and correlated risks where relevant.
  \item \textbf{Context and transport.} Declare team, role, lineup, opponent, coaching, data-vendor, and policy regimes; validate transport to the claimed context.
  \item \textbf{Feasible action set.} Reconstruct legal actions at the date and distinguish them from actions assumed to be commercially available.
  \item \textbf{Opportunity and selection.} Explain how realized lineups, rests, picks, contracts, and trades were selected; use identification, support checks, or sensitivity bounds.
  \item \textbf{Evaluation.} Match retrospective fit, temporal prediction, causal identification, policy simulation, constrained replay, or prospective trial to the claim being made.
  \item \textbf{Provenance and auditability.} Report source, access, version, transformations, code, and model cutoff. When data are private, state the public replication boundary and the internal audit mechanism.
\end{enumerate}

Targeted replication can resolve disagreements about definitions, leakage, or transport, and can give researchers a common benchmark. Reproducing every model is not necessary for the conceptual synthesis presented here. The case for a particular reproduction is strongest when it could change the interpretation of an important finding or settle a consequential uncertainty about its use.

\FloatBarrier
\section{Limitations}

The review draws together work from disciplines with different terminology and publication practices. Its iterative selection process helps connect those literatures but cannot establish that every relevant method paper has been found. Counts describe the retained corpus, and the qualitative map in Figure~\ref{fig:gapmap} should be read within that scope.

Publications also provide only a partial view of organizational practice. Teams have access to tracking, health, scouting, practice, negotiation, and internal objective-function information that researchers may never see. Limited public evidence for an integrated decision system does not establish that no team uses one. The review's conclusions concern published and publicly inspectable work.

Applying the six decision-readiness gates involves judgment about the claim a study actually makes. A sound descriptive study may have no reason to construct trades or estimate an intervention effect. The gates are intended to identify the evidence needed for a particular use, rather than produce a ranking of studies.

We assess the framework as a claim-matching and diagnostic structure by asking whether its gates follow from a decision made at a specified time, identify distinct failure modes, and yield coherent diagnoses across research streams. It has not been validated as a numerical scale or shown to improve team outcomes. Reliability as a scored instrument and prospective effects on decision quality require separate empirical investigation.

No game-level or draft outcomes under the new lottery were available at the cutoff. Asset-market analyses, however, require outcome-specific exposure dates and checks for anticipation around the announcement. CBA effects also remain difficult to isolate because restrictions phase in, teams anticipate exposure, and other changes affect payroll, participation, and competition at the same time. The suggested designs would need to address these sources of confounding.

Generalization beyond the NBA requires further work. Its small number of teams, concentrated talent, negotiated rules, and visible transactions make it a distinctive labor market. The framework may be useful in other leagues or workplaces, but its empirical conclusions cannot be assumed to transfer with it.

\section{Conclusion}

The reviewed literature provides substantial knowledge about player performance and several other components of basketball-operations decisions. The harder step is using that knowledge when the team, role, available alternatives, or rules differ from the setting in which an estimate was obtained. A forecast becomes more useful when those conditions are explicit and its uncertainty is carried into the comparison of actions.

The priority research agenda identified by this review is to test the links among these components. Contribution forecasts need evidence of role and team transport; lineup comparisons need workable rotations and availability assumptions; and acquisition models need realistic costs, rights, and alternatives. The six decision-readiness gates provide a way to specify what is missing in each case.

The 2023 CBA and the 3-2-1 Draft Lottery give this work practical urgency. They alter transaction options and draft incentives, creating settings in which researchers can preserve information before decisions and evaluate outcomes afterward. Progress will depend on whether the proposed links improve an actual comparison of feasible choices, as well as on the accuracy of the underlying estimates.

For studies that make a decision claim, the useful questions are specific: how could this estimate alter the comparison among feasible choices, what alternatives were available, and how would that comparison be evaluated? The framework is intended to help researchers answer them. Its usefulness in other constrained roster markets, and its effect on decisions within teams, remain open to empirical testing.

\section*{Data and materials availability}

This review synthesizes published research and publicly available official rule documents. The Supplementary Evidence Appendix accompanying this preprint documents the search and synthesis. The machine-readable retained-evidence map (\texttt{nba-decision-readiness-evidence-map.csv}) is openly available as a \href{https://github.com/yangzhou-tysportsanalytics/From-Metrics-to-Decisions-in-NBA-Analytics/blob/main/nba-decision-readiness-evidence-map.csv}{CSV on GitHub}. No player-level private health, tracking, scouting, or negotiation data were used, and ethical approval was not required.

\section*{Funding and competing interests}

The authors received no financial support for the research, authorship, or publication of this article and declare no competing interests.

\ifdefined\COMBINEDVERSION
\clearpage
\newgeometry{margin=0.72in}
\begingroup
\renewcommand{\normalsize}{\fontsize{10}{12}\selectfont}
\renewcommand{\small}{\fontsize{9}{11}\selectfont}
\renewcommand{\footnotesize}{\fontsize{8}{9.5}\selectfont}
\renewcommand{\scriptsize}{\fontsize{7}{8}\selectfont}
\normalsize
\setlength{\parindent}{1.1em}
\setlength{\parskip}{0.25em}
\setlength{\tabcolsep}{4.2pt}
\setlength{\LTpre}{0.45em}
\setlength{\LTpost}{0.8em}
\setlist[itemize]{leftmargin=1.4em,itemsep=0.15em,topsep=0.25em}
\setlist[enumerate]{leftmargin=1.6em,itemsep=0.15em,topsep=0.25em}
\fancyhf{}
\fancyhead[L]{\footnotesize Supplementary Evidence Appendix}
\fancyhead[R]{\footnotesize From Metrics to Decisions in NBA Analytics}
\fancyfoot[C]{\thepage}
\setcounter{table}{0}
\renewcommand{\theHtable}{S\arabic{table}}
\def\SUPPLEMENTBODYONLY{1}
\renewcommand{\thetable}{S\arabic{table}}

\begin{center}
  {\LARGE\bfseries\color{navy} Supplementary Evidence Appendix\par}
  \vspace{0.45em}
  {\large\bfseries From Metrics to Decisions in NBA Analytics:\par}
  {\large A Critical Integrative Review and Decision-Readiness Framework\par}
  \vspace{0.65em}
  {\normalsize Yang Zhou and Tianyu Guan\par}
  \vspace{0.35em}
  {\small Evidence cutoff: 1 September 2026}
\end{center}

\section*{S1. Purpose and review design}

This appendix records how evidence was selected and synthesized for the accompanying critical integrative review. It documents the links between NBA research streams, the limits of the claims drawn from them, and the judgments behind the qualitative evidence-gap map. The search was iterative rather than exhaustive. Counts describe the review's records; they are not estimates of research prevalence or study quality.

The evidence-discovery cutoff was 1 September 2026. Records were assembled from prior reviews, backward and forward citation tracing, targeted topic searches, publisher and DOI verification, and official NBA and NBA Players Association sources for rule states. Search functions and export capabilities differed across platforms, so searches were adapted to each source rather than represented as equivalent database runs. Bibliographic verification and targeted source and coding checks were completed through 17 September 2026. These did not expand the retained corpus or change the evidence cutoff of 1 September 2026.

\subsection*{S1.1 Discovery sources}

Sources searched or used for verification included scholarly web search; publisher and DOI pages; PubMed/MEDLINE and PubMed Central; arXiv, PMLR, SSRN, and MIT Sloan Sports Analytics Conference proceedings; NBER/IZA and institutional or author repositories; public code repositories linked to reports; and official NBA and NBA Players Association pages.

Where a platform supported structured fielded Boolean search, NBA or basketball terms were AND-combined with parenthesized OR synonyms in title, abstract, or keyword fields. Scholarly-web, venue, and repository searches used the same query families as broader keyword strings, followed by backward and forward citation chaining and title, DOI, and report-version verification. Field syntax was adapted to each platform rather than treated as identical across sources.

Representative query families were:

\begin{itemize}\small\ttfamily
  \item NBA (adjusted plus-minus OR RAPM OR player value) offense defense
  \item NBA (player role OR archetype OR profile) clustering longitudinal tracking
  \item NBA (shot selection OR shot quality OR shot making OR expected possession value)
  \item NBA lineup synergy interaction unseen lineup prediction network complementarity
  \item NBA substitution rotation minutes optimization fatigue lineup
  \item NBA injury availability workload prediction rest load management travel
  \item NBA draft prospect college combine scouting prediction mock consensus
  \item NBA draft position minutes retention survival sunk cost lottery tanking
  \item NBA salary contract compensation free agent valuation wage
  \item NBA trade acquisition roster team building optimization risk set candidate set
  \item site:nbpa.com/cba
  \item site:nba.com CBA 101 Second Apron
  \item NBA 3-2-1 Draft Lottery 2026 official
\end{itemize}

Title, author, DOI, cited-by, and version-of-record searches reconciled preprints, working papers, conference papers, and journal versions. When a publisher page was restricted, a lawful author manuscript or preprint was used when available.

\subsection*{S1.2 Eligibility and evidence roles}

Primary empirical studies, methodological studies demonstrated on NBA data, and full conference papers were eligible when they did at least one of the following:

\begin{enumerate}
  \item estimated a decision-relevant state;
  \item evaluated uncertainty or transport;
  \item modeled an intervention, candidate action, or constrained choice; or
  \item exposed a material data, selection, market, or institutional boundary for basketball operations.
\end{enumerate}

Reviews were retained for prior-coverage comparison and terminology. Official NBA and NBA--NBPA documents were retained only to define rule and policy states, not as outcome evidence. Broader basketball or multisport work was retained only when its method or evidence directly informed an NBA decision interface.

Records were excluded when they were duplicates or superseded versions, had no NBA empirical or direct decision-interface contribution, repeated a sample without adding a distinct estimand, addressed betting/fan/media topics outside the stated scope, focused on a clinical treatment without an availability or operations interface, or remained an abstract-only lead not required to delimit the manuscript's scope. Retained abstract-only records are marked and do not carry decisive field-level findings.

\subsection*{S1.3 Record accounting and deduplication}

The five discovery workstreams organized the search and extraction process. Their reports were subsequently grouped into seven research streams for the synthesis.

\begin{table}[ht]
\centering
\caption{Discovery records by source evidence workstream.}
\begin{tabular}{lr}
\toprule
\textbf{Source evidence workstream} & \textbf{Rows} \\
\midrule
Player value, role, and on-court action & 21 \\
Synergy and lineup decisions & 27 \\
Availability, injury, and workload & 30 \\
Draft, development, and lottery & 25 \\
Contracts, acquisition, and roster construction & 29 \\
\midrule
Total discovery rows & 132 \\
\bottomrule
\end{tabular}
\end{table}

Deduplication used, in order, a normalized DOI; a normalized title when no DOI was available; and manual inspection of authors, year, and report/version identity. Working-paper and journal manifestations of the same study were treated as one report. Five duplicate report pairs were reconciled across workstreams, leaving 127 unique workstream reports.

\begin{table}[ht]
\centering
\caption{Retained evidence accounting.}
\begin{tabular}{lr@{\qquad}lr}
\toprule
\textbf{NBA synthesis status} & \textbf{Reports} & \textbf{Manuscript evidence category} & \textbf{Reports} \\
\midrule
Full text inspected & 50 & Primary empirical study & 21 \\
Abstract/metadata only & 5 & Methodological study & 24 \\
Official rule document & 3 & Full conference paper & 2 \\
 & & Preprint & 2 \\
 & & Review & 6 \\
 & & Official rule & 3 \\
\midrule
NBA synthesis corpus & 58 & NBA synthesis corpus & 58 \\
Framework foundations & 8 & Combined citation map & 66 \\
\bottomrule
\end{tabular}
\end{table}

The right-hand categories are mutually exclusive coding categories for manuscript accounting, not a universal publication taxonomy. When a source could carry multiple labels, it is counted under the category that describes how it enters this review.

Of the 127 unique workstream reports, 51 were retained and the remaining 76 remain discovery or background records. Seven additional foundations bring the cited NBA synthesis corpus to 58 reports: the six prior reviews retained for scope comparison (Sarlis and Tjortjis, 2020; Terner and Franks, 2021; Huyghe et al., 2022; Kovalchik, 2023; Zhou and Li, 2024; Chen et al., 2025) and Kubatko et al. (2007) as the NBA statistical foundation. Eight non-NBA methodological sources ground the decision-readiness framework and are excluded from NBA evidence counts and Figure 4 maturity judgments.

\section*{S2. Extraction and synthesis}

The workstream tables record each report's bibliographic identity, publication and access status, seasons, data sources, observation unit, estimand, method, and validation. They also distinguish findings supported by the source from inferences made in this review, and record uncertainty, limitations, decision tasks, action and opportunity boundaries, provenance, and public replication feasibility.

The harmonized evidence map records the furthest claim level directly used by the manuscript: measurement, prediction, causal or counterfactual estimation, constrained prescription, or prospective evaluation. It separately records point-in-time validity, uncertainty, context portability, action feasibility, opportunity-set observability, evaluation, and reproducibility or auditability. Fields use \emph{yes}, \emph{partial}, \emph{no}, \emph{not applicable}, or \emph{not reported}; they are not summed into a score. In the \href{https://github.com/yangzhou-tysportsanalytics/From-Metrics-to-Decisions-in-NBA-Analytics/blob/main/nba-decision-readiness-evidence-map.csv}{GitHub-hosted CSV}, the \texttt{source\_table} field identifies an internal discovery workstream rather than a separate evidence source. Its \texttt{evidence\_id} and \texttt{citation\_key} fields link retained records to Section~S4 and the reference list; the remaining fields record evidence role, access, claim level, diagnostic assessments, manuscript use, and interpretive notes.

The combined \texttt{reproducibility\_auditability} field is interpreted alongside access status and the evidence notes. For AV01, AV02, AV06, and AV10, \emph{no} denotes restricted public reproducibility, not the absence of internal auditability. AV02 describes an internally audited record system.

We recorded the furthest analytical claim implemented and used in this review, and assessed its validation and decision readiness separately. We then considered the readiness gates relevant to its claim and the evidence still needed for the proposed use. Descriptive work was assessed on its own terms. Abstract-only records established the presence of a topic or method, but were not used to support numerical effects or mature interfaces. Official documents established rules and dates, without establishing policy effectiveness.

Both authors worked jointly on screening, extraction, and coding. During manuscript preparation, both authors independently cross-checked the source claims and the extracted information for all 66 records included in Table~S4, as well as all Figure 4 cell ratings. Discrepancies were resolved through joint review, and the final coding was confirmed by both authors. Evidence extraction was domain-specific and then harmonized for this review. No formal estimate of inter-rater agreement is reported. No paper reproduction or PRISMA flow is claimed.

\section*{S3. Figure 4 maturity definitions and audit trail}

The unit of classification is a specified research interface at one claim level. Table~\ref{tab:maturity-audit} states the target of each cell and the evidence or boundary supporting its rating. Measurement concerns the stated representation or construct; prediction concerns outcomes in a declared future or held-out sample; causal/counterfactual estimation concerns intervention effects or model-based alternative outcomes; constrained prescription concerns comparison of candidate actions under an objective and explicit constraints; prospective evaluation concerns an implemented decision rule and a credible comparator. Model-based counterfactuals are not treated as identified causal effects.

Coding first distinguishes direct implementations from adjacent evidence, then considers the accumulation of distinct full-text contributions and validation appropriate to the target. Report versions of one study do not provide independent support; distinct contributions can share NBA data and do not automatically constitute external replication. Measurement is assessed through construct, measurement, or robustness evidence; prediction through relevant held-out or temporal evaluation; causal/counterfactual work through its identification or structural assumptions and sensitivity; action models through candidate comparisons and constraint coverage; and deployment through the implemented comparison. Requirements from later claim levels are not imposed on earlier ones.

\begin{itemize}
  \item \textbf{Developed:} Multiple distinct full-text contributions form an identifiable research line, with established implementations and repeated empirical support appropriate to the specified target and claim level. Remaining limitations do not reduce the evidence to isolated demonstrations. Developed does not mean free of bias or decision-ready.
  \item \textbf{Emerging:} Substantive direct implementation extends beyond an isolated illustrative proof of concept, but accumulation or claim-appropriate validation remains too limited for Developed. Identification, transport, calibration, or constraint coverage may remain unresolved; their relevance depends on the target.
  \item \textbf{Thin:} Direct evidence is absent or confined to isolated illustrations; other anchors are adjacent, diagnostic, or conceptual. Later-stage gaps alone do not imply Thin at an earlier claim level.
\end{itemize}

The rules are applied in that order: an interface that does not meet Developed is assessed for substantive direct work before being assigned Emerging or Thin. Boundary rationales identify the limiting evidence rather than require a universal end-to-end system. No numerical score or fixed publication-count threshold is used. Public literature includes inspectable reports based on restricted data; lack of access to raw data is recorded separately and is not by itself a reason to lower maturity. Abstract-only reports establish topic presence, and official rules establish legal states, but neither establishes a mature empirical interface. Evidence IDs resolve to Section~S4. An anchor may document a limitation rather than supply positive evidence at the cell's claim level; such use does not upgrade the report's own claim.

\paragraph{Targeted source checks.} The following checks clarify important boundaries in Tables~S3 and S4:
\begin{itemize}
  \item PV02, \citet{Sill2010}, pp.~3--6: the March--April 2009 holdout uses coefficients fitted earlier, but predictions condition on realized lineups and possessions. PV03, \citet{DeshpandeJensen2016}, Sections~1 and 7: player impact is explicitly retrospective, not a forecast. PV03 therefore supports measurement; the retained direct prediction evidence is insufficient for Developed.
  \item RO01, \citet{SkinnerGuy2015}, Section~5 and Methods: the demonstration uses 780 hand-coded sequences from three playoff games, with limited cross-lineup checks. A tracking-oriented method is not itself evidence of large-scale tracking validation.
  \item SY03, \citet{Kuehn2017}: the retained extraction record and publisher description support model-based candidate trades. The \texttt{opportunity\_set} code is \emph{partial}, because the retained record does not establish actual market availability.
  \item AV01/AV02 support injury and participation ascertainment. The Availability measurement rating also covers workload representation (AV06); its limited validation as an indicator of capacity should not be read as a judgment that injury surveillance is immature.
  \item AV10, \citet{Herzog2026}, Sections~2.3 and 3.1: no statistically significant regular-season injury-risk differences were found across rest/load-management groups for in-game injuries causing at least two consecutive missed games. This observational result does not establish an individual causal effect.
  \item DR01, \citet{Berri2011}, Sections~3--4 and Tables~3 and 5--6: retrospective regressions relate pre-draft characteristics to selection and later production; no held-out predictive evaluation is reported. DR03, \citet{FisherMontague2025}, evaluates dated selection-order forecasts. This supports Emerging for draft-order/consensus prediction, without establishing prediction of player productivity or development effects.
  \item CR05, \citet{Brill2023}, Sections~2--3: a modeled lineup outcome is used to rank acquisition candidates under a salary filter, without reported held-out forecast or recommendation validation. CR02/CR04 supply compensation and accounting context. Emerging denotes direct modeling with limited validation.
\end{itemize}

\small
\begin{longtable}{P{2.25cm}P{2.00cm}P{1.60cm}P{2.45cm}P{5.45cm}}
\caption{Audit trail for the seven-by-five qualitative maturity map in Figure 4. Each rationale specifies the target, supporting evidence, and relevant boundary. Evidence IDs are representative anchors rather than citation counts.}\label{tab:maturity-audit}\\
\toprule
\textbf{Stream} & \textbf{Claim level} & \textbf{Rating} & \textbf{Evidence IDs} & \textbf{Target, evidence, and boundary} \\
\midrule
\endfirsthead
\multicolumn{5}{l}{\footnotesize\textit{Table \thetable\ continued from previous page}}\\
\toprule
\textbf{Stream} & \textbf{Claim level} & \textbf{Rating} & \textbf{Evidence IDs} & \textbf{Target, evidence, and boundary} \\
\midrule
\endhead
\midrule
\multicolumn{5}{r}{\footnotesize\textit{Continued on next page}}\\
\endfoot
\bottomrule
\endlastfoot
On-court action & Measurement & Developed & OA01, OA03, OA04, OA07 & Observed possession, spatial behavior, and action value: accounting, defensive attribution, and action-value representations form established measurement lines. \\
On-court action & Prediction & Developed & OA05, OA06 & Downstream possession value: possession and transition models provide empirical prediction and validation with explicit state dynamics. \\
On-court action & Causal / counterfactual & Emerging & OA02, OA06, OA07 & Alternative micro-action outcomes: explicit modeled comparisons extend beyond conceptual proposals, but unchosen-action support and strategic response limit causal interpretation. \\
On-court action & Constrained prescription & Emerging & OA02, OA06, OA07 & Pass and shot choice: models enumerate and compare actions; capability, teachability, and opponent adaptation remain incompletely represented. \\
On-court action & Prospective evaluation & Thin & OA05, OA06, OA07 & Implemented coaching rules: the retained anchors evaluate models, not a prospectively deployed coaching decision against a credible comparator. \\
Player value & Measurement & Developed & PV01, PV03, PV04, PV05, PV06 & Observed-context contribution: multiple impact estimators and metric audits examine attribution, stability, and ranking uncertainty. This target does not require team-switch validation. \\
Player value & Prediction & Emerging & PV02, PV03 & Held-out game margins: PV02 validates conditional predictions using realized lineups and possessions; PV03 is retrospective measurement. Direct predictive accumulation in the retained corpus is insufficient for Developed. \\
Player value & Causal / counterfactual & Emerging & PV05, PV06 & Teammate effects on impact metrics: PV06 implements FE/IV analyses; PV05 supplies a diagnostic audit, not causal evidence. Identification assumptions and portable player effects remain unresolved. \\
Player value & Constrained prescription & Thin & PV01, PV03, PV06 & Deployment or acquisition choice: the anchors estimate or audit player contribution without comparing implementable actions under an objective and constraints. \\
Player value & Prospective evaluation & Thin & PV02, PV03 & Implemented player-value rules: PV02 supplies conditional predictive testing and PV03 retrospective measurement; neither prospectively evaluates their use in team decisions. \\
Role / capability & Measurement & Developed & RO01, RO02 & Role representation: skill features and multi-season archetypes provide empirically studied representations and stability checks beyond nominal positions. \\
Role / capability & Prediction & Emerging & RO01, RO02 & Performance from role features: RO01 tests limited cross-lineup skill transfer using hand-coded games; RO02 supplies representation stability. Validation after team or role changes is sparse. \\
Role / capability & Causal / counterfactual & Thin & RO01, RO02 & Role-reassignment outcomes: neither representation anchor directly estimates an intervention or models outcomes under alternative assignments. \\
Role / capability & Constrained prescription & Thin & RO01, RO02 & Role or minutes assignment: representations suggest applications but do not compare implementable assignments under a declared objective and constraints. \\
Role / capability & Prospective evaluation & Thin & RO01, RO02 & Representation-guided reassignment: no retained study prospectively evaluates an implemented role-assignment rule. \\
Synergy / lineup & Measurement & Developed & SY01, SY02, SY03, SY04, SY06 & Observed teammate interaction: pair effects, spillovers, higher-order decompositions, and networks constitute an established measurement literature. \\
Synergy / lineup & Prediction & Emerging & SY05, RO01 & Unseen-lineup performance: SY05 uses time-ordered prediction; RO01 provides limited within-series cross-lineup checks. Validation remains sparse and conditioned on lineups coaches used. \\
Synergy / lineup & Causal / counterfactual & Emerging & SY01, SY02, SY03 & Alternative teammate configurations: structural production and possession models provide counterfactual comparisons. Strategic selection and structural assumptions prevent treating these as identified deployment effects. \\
Synergy / lineup & Constrained prescription & Emerging & SY01, SY03, SY05 & Lineup or roster choice: simulated trades and candidate screening compare actions, unlike representation alone; complete rotation constraints and market access remain unvalidated. \\
Synergy / lineup & Prospective evaluation & Thin & SY01, SY05 & Implemented lineup policies: no retained system prospectively evaluates a substitution, minutes, or lineup rule against a credible alternative. \\
Availability & Measurement & Emerging & AV01, AV02, AV06 & Participation and workload representation: AV01/AV02 support injury and participation ascertainment. Emerging concerns limited validation of workload measures as capacity indicators (AV06), not event surveillance or restricted data access itself. \\
Availability & Prediction & Emerging & AV04, AV05 & Future injury or absence risk: direct risk models exist, but label validity, temporal validation, and calibration remain limited. \\
Availability & Causal / counterfactual & Emerging & AV08, AV09, AV10 & Rest effects on health or performance: empirical intervention contrasts and heterogeneity analyses exist, but confounding by indication and interference remain unresolved. \\
Availability & Constrained prescription & Thin & AV06, AV09, AV10 & Workload or rest choice: monitoring and effect estimates do not directly compare feasible policies under an objective incorporating health, replacement performance, and competitive cost. \\
Availability & Prospective evaluation & Thin & AV04, AV10 & Implemented availability rules: risk prediction and retrospective rest analysis do not prospectively compare a deployed policy with a credible alternative. \\
Draft / development & Measurement & Emerging & DR02, DR03, DR05 & Prospect signals: athletic tests, public ranks, and allocated opportunity are measured, but construct validation separating ability, belief, and development opportunity remains limited. \\
Draft / development & Prediction & Emerging & DR01, DR03 & Draft-order/consensus prediction: DR03 evaluates dated forecasts against actual selections; DR01 reports retrospective regressions without held-out evaluation. Repeated direct predictive validation is insufficient for Developed; future productivity and development effects remain outside this target. \\
Draft / development & Causal / counterfactual & Emerging & DR04, DR06, DR07, DR08 & Information and lottery-incentive effects: quasi-experimental and regime comparisons address alternatives, but concurrent changes, opportunity, and behavioral interpretation limit identification. \\
Draft / development & Constrained prescription & Thin & DR01, DR03, DR09 & Pick or prospect acquisition: prediction and rule anchors inform a choice but do not directly compare priced, team-specific candidate actions under constraints. \\
Draft / development & Prospective evaluation & Thin & DR08, DR09 & Implemented draft rules: no retained study evaluates a prospectively deployed team decision rule; the new lottery also had no treated season by the cutoff. \\
Contract / roster & Measurement & Emerging & CR02, CR03, CR04 & Performance-to-compensation value: contract outcomes, consumer value, and realized ROI have direct empirical measures, but different targets and limited validation of value conversion prevent an established common interface. \\
Contract / roster & Prediction & Emerging & CR02, CR04, CR05 & Prediction-based acquisition ranking: CR05 ranks candidates using modeled lineup outcomes, but reports no held-out forecast or recommendation validation. CR02/CR04 provide compensation and accounting context, not forecast validation. \\
Contract / roster & Causal / counterfactual & Emerging & CR02, CR03 & Compensation under alternative rules: CR03 models a pay counterfactual; CR02 supplies observed-contract context. Assumptions about valuation and selection limit interpretation. \\
Contract / roster & Constrained prescription & Emerging & CR05, CR06, CR10, CR11, CR12 & Acquisition or roster choice: models compare candidates under explicit objectives and simplified constraints. Rule documents define missing legal requirements; full CBA and market coverage remain unvalidated. \\
Contract / roster & Prospective evaluation & Thin & CR06, CR10 & Implemented roster rules: retrospective cases and stylized simulations do not establish prospective NBA execution or decision improvement. \\
\end{longtable}

\normalsize

\section*{S4. Retained-evidence crosswalk}

The table links each citation to its role in the review and the main limitation on its use in a decision. It distinguishes the 58 NBA reports from the eight general framework sources. The furthest claim level describes the implemented analysis used here, not its overall quality or validation status. A prediction-oriented model is distinguished from held-out predictive validation in the evidence notes and maturity rationales. A descriptive study can be sound without constructing a trade; a prescriptive claim needs evidence about feasibility, opportunity, uncertainty, and evaluation.

\scriptsize
\begin{longtable}{P{1.00cm}P{2.85cm}P{2.45cm}P{2.15cm}P{4.77cm}}
\caption{Retained-evidence crosswalk. The 58 NBA synthesis reports and eight framework-foundation sources are shown separately by the Stream/type column. ``Furthest claim'' records the highest analytical reach used in the manuscript, not an overall quality rating.}\label{tab:evidence-crosswalk}\\
\toprule
\textbf{ID} & \textbf{Source} & \textbf{Stream / type} & \textbf{Furthest claim} & \textbf{Decision interface and principal boundary} \\
\midrule
\endfirsthead
\multicolumn{5}{l}{\footnotesize\textit{Table \thetable\ continued from previous page}}\\
\toprule
\textbf{ID} & \textbf{Source} & \textbf{Stream / type} & \textbf{Furthest claim} & \textbf{Decision interface and principal boundary} \\
\midrule
\endhead
\midrule
\multicolumn{5}{r}{\footnotesize\textit{Continued on next page}}\\
\endfoot
\bottomrule
\endlastfoot
REV01 & \citet{Sarlis2020} & cross-stream review; review & not applicable & prior-review positioning. Reviews player and team evaluation; used for scope comparison, not outcome evidence. \\
REV02 & \citet{Terner2021} & cross-stream review; review & not applicable & prior-review positioning. Methodological review of basketball player and team performance. \\
REV03 & \citet{Huyghe2022} & cross-stream review; review & not applicable & prior-review positioning. NBA gameplay-performance systematic review; salaries and health/injury are outside its stated scope. \\
REV04 & \citet{Kovalchik2023} & cross-stream review; review & not applicable & prior-review positioning and data-access boundary. Cross-sport tracking-data review used for method and access framing. \\
REV05 & \citet{Chen2025} & cross-stream review; review; abstract/metadata only & not applicable & prior-review positioning. Only high-level abstract-supported coverage claims are retained; not used for detailed findings. \\
REV06 & \citet{ZhouLi2024} & cross-stream review; review & not applicable & prior-review positioning. Full text supports the reported dimension--granularity--task--stakeholder-question organizing structure and its 32-question agenda. \\
OA01 & \citet{Kubatko2007} & on-court action; cross-stream foundation; methodological study & measurement & observation-to-state. Foundational possession accounting and basketball statistics; not a modern decision-support validation. \\
OA02 & \citet{Skinner2012} & on-court action; methodological study & constrained prescription & state-to-action. Formalizes shot-selection tradeoffs; empirical intervention and opponent-response validation remain limited. \\
OA03 & \citet{Miller2014} & on-court action; full conference paper & measurement & observation-to-state. Spatial shot-intensity model using tracking-era data; measurement rather than a tested coaching intervention. \\
OA04 & \citet{FranksDefense2015} & on-court action; methodological study & measurement & observation-to-state. Measures defensive spatial responsibility and skill; role and assignment transport are not prospectively tested. \\
OA05 & \citet{Cervone2016} & on-court action; methodological study & prediction & state-to-outcome. Possession-value prediction supports action comparison but does not identify unchosen-action effects. \\
OA06 & \citet{Sandholtz2020} & on-court action; methodological study & constrained prescription & state-to-action. Transition model evaluates observed on-court decisions; uncommon or unchosen actions and deployment remain bounded. \\
OA07 & \citet{Jutamulia2025} & on-court action; methodological study & measurement & state-to-action. Expected action value enumerates immediate pass and shot alternatives and measures decision skill; the comparisons are predictive rather than causal or prospectively evaluated. \\
PV01 & \citet{FearnheadTaylor2011} & player value; methodological study & measurement & observation-to-state. Posterior offense/defense contribution in observed lineups; future team-switch portability is not tested. \\
PV02 & \citet{Sill2010} & player value; full conference paper & prediction & state-to-outcome. Conditional held-out game-margin prediction; coefficients fitted before the March-April 2009 test games, but realized lineups and possession counts enter predictions (pp. 3-6). No prospective decision evaluation. \\
PV03 & \citet{DeshpandeJensen2016} & player value; methodological study & measurement & observation-to-state. Retrospective, context-dependent contribution to winning, explicitly not future player performance (Sections 1 and 7). Internal win-probability estimation does not validate a player-contribution forecast. \\
PV04 & \citet{Barrientos2023} & player value; methodological study & measurement & state-to-ranking. Makes uncertainty in rankings explicit; not a team-specific choice evaluation. \\
PV05 & \citet{FranksMeta2016} & player value; methodological study & measurement & metric-to-state audit. Assesses discrimination, stability, and independence of sports metrics; supplies diagnostic context rather than direct causal evidence. \\
PV06 & \citet{Ghimire2020} & player value; primary empirical study & causal or counterfactual estimation & teammate context-to-impact metric. FE/IV specifications examine teammate effects on RPM; instrument validity and selection remain assumptions, and a portable player effect is not identified. \\
RO01 & \citet{SkinnerGuy2015} & role / capability; methodological study & prediction & observation-to-state. Tracking-oriented skill model demonstrated on 780 hand-coded sequences from three 2011 playoff games; limited cross-lineup checks, not broad team-switch validation (Section 5 and Methods). \\
RO02 & \citet{Muniz2022} & role / capability; methodological study & measurement & observation-to-state. Recovers stable player archetypes across seasons; clusters do not identify causal role assignments. \\
SY01 & \citet{Maymin2013} & synergy / lineup; methodological study & constrained prescription & state-to-action. Pairwise chemistry and simulated trades; lineup selection and market availability are not identified. \\
SY02 & \citet{Arcidiacono2017} & synergy / lineup; primary empirical study & causal or counterfactual estimation & teammate configuration-to-production. Structural spillover estimates and alternative-team comparisons depend on conditional exogeneity; substitution robustness does not identify a deployed lineup effect. \\
SY03 & \citet{Kuehn2017} & synergy / lineup; methodological study & constrained prescription & state-to-action. Lineup-specific marginal values and model-based trade comparisons depend on the possession model; one-season fit, selection, and unclear provenance restrict deployment claims. Candidate alternatives are enumerated within the model; the retained evidence record does not establish actual market availability. \\
SY04 & \citet{DevlinUminsky2020} & synergy / lineup; methodological study & measurement & observation-to-state. Decomposes observed lineup performance; no predictive or intervention validation. \\
SY05 & \citet{Martonosi2023} & synergy / lineup; methodological study & prediction & state-to-outcome. Time-ordered unseen-lineup prediction, but evaluation is conditioned on lineups coaches later used. \\
SY06 & \citet{Fewell2012} & synergy / lineup; primary empirical study & measurement & observation-to-state. Strategic passing networks describe team process; they do not validate a lineup or tactical intervention. \\
AV01 & \citet{Drakos2010} & availability; primary empirical study & measurement & observation-to-state. Internal surveillance measures diagnosed injury burden and participation loss; it does not measure complete workload capacity. The no code concerns restricted public reproducibility; it does not establish the absence of internal auditability. \\
AV02 & \citet{Mack2019} & availability; methodological study & measurement & data-to-state governance. Standardizes internal injury, illness, and participation records; raw-data access is a replication boundary, not itself a maturity penalty. The no code concerns restricted public reproducibility; the source describes an internally audited record system. \\
AV03 & \citet{Teramoto2017} & availability; primary empirical study & measurement & exposure-to-state. Associates schedule context with in-game injury frequency; not a causal rest effect. \\
AV04 & \citet{Lewis2018} & availability; primary empirical study & prediction & state-to-outcome. Pregame public-data risk model; no temporal external validation, intervention model, or calibrated decision utility. \\
AV05 & \citet{Cohan2021} & availability; methodological study & prediction & state-to-outcome. Random-split injury classifier with extreme imbalance; temporal and decision validation are absent. \\
AV06 & \citet{Russell2021} & availability; primary empirical study & measurement & observation-to-state. Single-team workload study compares practice and game load; cross-team validation and physiological-capacity measurement remain limited. The no code concerns restricted public reproducibility; it does not establish the absence of internal auditability. \\
AV07 & \citet{Charest2021} & availability; primary empirical study & measurement & exposure-to-state. Travel and time-zone associations are reproducible proxies, not player-level health effects. \\
AV08 & \citet{Belk2017} & availability; primary empirical study & causal or counterfactual estimation & action-to-outcome. Matched retrospective rest study; small sample and selection prevent a strong causal conclusion. \\
AV09 & \citet{NakamuraSakai2024} & availability; preprint & causal or counterfactual estimation & action-to-outcome. Heterogeneous rest-effect model; causal interpretation depends on unconfoundedness and no interference, and external validation remains unresolved. \\
AV10 & \citet{Herzog2026} & availability; primary empirical study & causal or counterfactual estimation & action-to-outcome. No statistically significant regular-season injury-risk differences across rest/load-management groups for in-game injuries causing at least two consecutive missed games (Sections 2.3 and 3.1); observational comparison, not an identified individual rest effect. The no code concerns restricted public reproducibility; it does not establish the absence of internal auditability. \\
DR01 & \citet{Berri2011} & draft / development; primary empirical study & prediction & state-to-outcome. Retrospective regressions relate pre-draft characteristics to draft position and later NBA production; no held-out predictive evaluation is reported (Sections 3-4, Tables 3 and 5-6). The prediction label records the modeling target, not validated forecast performance. \\
DR02 & \citet{BergerDaumann2021} & draft / development; primary empirical study & measurement & signal-to-market belief. Combine athleticism predicts selection more than later observed performance; missing tests and opportunity are selected. \\
DR03 & \citet{FisherMontague2025} & draft / development; methodological study & prediction & public belief-to-selection. Date-indexed public mock aggregation evaluates draft-order forecasts, not future NBA productivity or development effects; the historical corpus is not released (Sections 5-6). \\
DR04 & \citet{IchniowskiPreston2017} & draft / development; primary empirical study & causal or counterfactual estimation & information-to-market belief. Uses pre-tournament mock rank as a dated prior; private boards and later opportunity remain unobserved. \\
DR05 & \citet{StawHoang1995} & draft / development; primary empirical study & measurement & selection-to-opportunity. Shows draft order is associated with later opportunity and survival; does not cleanly identify a sunk-cost mechanism. \\
DR06 & \citet{TaylorTrogdon2002} & draft / development; primary empirical study & causal or counterfactual estimation & rule-to-behavior. Three-regime comparison supports incentive response but not directly observed intent. \\
DR07 & \citet{Price2010} & draft / development; primary empirical study & causal or counterfactual estimation & rule-to-behavior. Long-panel extension strengthens historical generalization while team intent and mechanism remain latent. \\
DR08 & \citet{SchmidtLottery2024} & draft / development; primary empirical study & causal or counterfactual estimation & rule-to-roster-use. One-season-before/one-season-after portfolio analysis is informative but insufficient for a standalone causal policy claim. \\
DR09 & \citet{NBA2026Lottery} & draft / development; official rule & not applicable & rule-state definition. Authoritative treatment definition only; zero completed treated seasons at the evidence cutoff. \\
CR01 & \citet{Ehrlich2019} & contract / roster; primary empirical study; abstract/metadata only & measurement & performance-to-market price. Abstract-level evidence only; used to identify the offense-defense wage question, not a decisive effect estimate. \\
CR02 & \citet{Wen2023} & contract / roster; primary empirical study & measurement & performance-to-contract outcome. Observed contract salary, duration, and performance have specification checks but no predictive holdout; rejected terms and causal contract effects are not observed. \\
CR03 & \citet{Kaplan2024} & contract / roster; primary empirical study & causal or counterfactual estimation & rule-to-compensation distribution. Counterfactual pay distribution under min/max constraints; league-wide consumer value is not team-specific contract value. \\
CR04 & \citet{Lautier2025} & contract / roster; methodological study & measurement & performance-and-salary-to-ROI. Realized salary ROI is an accounting allocation with internal checks, not a signing-time forecast or independently validated marginal contract value. \\
CR05 & \citet{Brill2023} & contract / roster; methodological study & constrained prescription & state-to-acquisition candidate. Prediction-based candidate ranking uses modeled lineup outcomes and a simplified salary filter; no held-out forecast or recommendation validation is reported (Sections 2-3). Legal and market availability remain unverified. \\
CR06 & \citet{MunizFlamand2023} & contract / roster; methodological study & constrained prescription & state-to-roster action. Mixed-integer roster construction demonstrates budget-feasible alternatives but omits the full CBA and counterpart choice. \\
CR07 & \citet{Ke2024} & contract / roster; methodological study; abstract/metadata only & constrained prescription & state-to-roster action. Abstract-only record; does not carry a decisive field-level maturity judgment. \\
CR08 & \citet{MayminGM2017} & contract / roster; methodological study; abstract/metadata only & constrained prescription & state-to-personnel action. Abstract-only record; draft, trade, and free-agency system is noted without treating reported performance as validated deployment. \\
CR09 & \citet{SchmidtTeamBuilding2021} & contract / roster; primary empirical study; abstract/metadata only & measurement & market state-to-team-building risk. Abstract-only record; supports the existence of a risk/uncertainty question, not a decision-readiness conclusion. \\
CR10 & \citet{Zhang2026} & contract / roster; preprint & constrained prescription & state-to-dynamic roster action. Rolling-horizon stochastic optimization is evaluated in a stylized simulation, not an audited NBA transaction environment. \\
CR11 & \citet{NBA2023CBA} & contract / roster; official rule & not applicable & rule-state definition. Authoritative agreement defines legal states; it is not evidence that a provision caused an outcome. \\
CR12 & \citet{NBA2024CBA101} & contract / roster; official rule & not applicable & rule-state explainer. Official explainer supplements but does not replace the signed CBA. \\
FF01 & \citet{Tashman2000} & framework foundation; review & not applicable & point-in-time and rolling-origin validation foundation. Forecast-evaluation foundation; excluded from the NBA synthesis corpus and Figure 4 ratings. \\
FF02 & \citet{PearlBareinboim2014} & framework foundation; methodological study & not applicable & transportability and external-validity foundation. Transportability foundation; excluded from the NBA synthesis corpus and Figure 4 ratings. \\
FF03 & \citet{Manski2003} & framework foundation; methodological study & not applicable & partial-identification foundation. Partial-identification foundation for hidden opportunity sets; excluded from the NBA synthesis corpus and Figure 4 ratings. \\
FF04 & \citet{SwaminathanJoachims2015} & framework foundation; full conference paper & not applicable & off-policy evaluation foundation. Logged-bandit and counterfactual-risk foundation; excluded from the NBA synthesis corpus and Figure 4 ratings. \\
FF05 & \citet{Berger1985} & framework foundation; methodological study & not applicable & statistical decision-theory foundation. Decision-theory foundation for actions, uncertainty, utilities, and losses; excluded from the NBA synthesis corpus and Figure 4 ratings. \\
FF06 & \citet{Howard1966} & framework foundation; methodological study & not applicable & value-of-information foundation. Value-of-information foundation; excluded from the NBA synthesis corpus and Figure 4 ratings. \\
FF07 & \citet{BertsimasBrownCaramanis2011} & framework foundation; review & not applicable & robust-optimization foundation. Robust-optimization foundation; excluded from the NBA synthesis corpus and Figure 4 ratings. \\
FF08 & \citet{BoydVandenberghe2004} & framework foundation; methodological study & not applicable & constrained-optimization foundation. Constrained-optimization foundation; excluded from the NBA synthesis corpus and Figure 4 ratings. \\
\end{longtable}

\normalsize

\section*{S5. Limitations of the evidence record}

\begin{enumerate}
  \item Discovery was iterative and source capabilities were unequal; the corpus is not an exhaustive enumeration of NBA analytics.
  \item The five domain tables used different native schemas. Harmonization improves comparability but cannot recover information a source workstream did not record, including a complete set of individual exclusion reasons for the 76 non-retained reports.
  \item Five retained records are abstract/metadata only and are restricted to positioning or scope claims.
  \item Some studies use commercial tracking, internal medical data, unreleased historical collections, or unobserved bargaining and opportunity sets. Access to the paper does not imply that its analysis can be publicly replicated.
  \item The maturity map is a structured qualitative synthesis, not a validated numerical scale. Ratings concern public research interfaces rather than individual paper quality.
  \item The review applies the framework across research streams. It does not test whether adopting the framework improves team outcomes.
\end{enumerate}

The machine-readable companion file (\texttt{nba-decision-readiness-evidence-map.csv}) is openly available as a \href{https://github.com/yangzhou-tysportsanalytics/From-Metrics-to-Decisions-in-NBA-Analytics/blob/main/nba-decision-readiness-evidence-map.csv}{CSV on GitHub}. Domain-level evidence tables preserve the more detailed extraction fields used to produce this appendix.

\clearpage
\endgroup
\restoregeometry
\normalsize
\fancyhf{}
\fancyhead[L]{\small From Metrics to Decisions in NBA Analytics}
\fancyhead[R]{\small Critical integrative review}
\fancyfoot[C]{\thepage}
\fi

\end{document}